\documentclass[10pt,twocolumn]{article}          
\usepackage{latex8}                              
\usepackage{psfig}
\usepackage{times}                               

\begin{document}

\title{Active Virtual Network Management Prediction: Complexity as a
  Framework for Prediction, Optimization, and Assurance} 

\author{Stephen F. Bush\\ 
General Electric Corporate Research and Development\\
1 Research Circle\\
Niskayuna, NY 12309\\
bushsf@crd.ge.com\\
http://www.crd.ge.com/\~{}bushsf/ftn}

\maketitle
\thispagestyle{empty}

\begin{abstract}
Research into active networking has provided the incentive to re-visit
what has traditionally been classified as distinct properties and
characteristics of information transfer such as protocol versus
service; at a more fundamental level this paper considers the blending
of computation and communication by means of complexity. The specific
service examined in this paper is network self-prediction enabled by
Active Virtual Network Management
Prediction. Computation/communication is analyzed via Kolmogorov
Complexity. The result is a mechanism to understand and improve the
performance of active networking and Active Virtual Network Management
Prediction in particular. The Active Virtual Network Management
Prediction mechanism allows information, in various states of
algorithmic and static form, to be transported in the service of
prediction for network management. The results are generally
applicable to algorithmic transmission of information. Kolmogorov
Complexity is used and experimentally validated as a theory describing
the relationship among algorithmic compression, complexity, and
prediction accuracy within an active network. Finally, the paper
concludes with a complexity-based framework for Information Assurance
that attempts to take a holistic view of vulnerability analysis.
\end{abstract}

\begin{figure}[ht]
  \centerline{\psfig{file=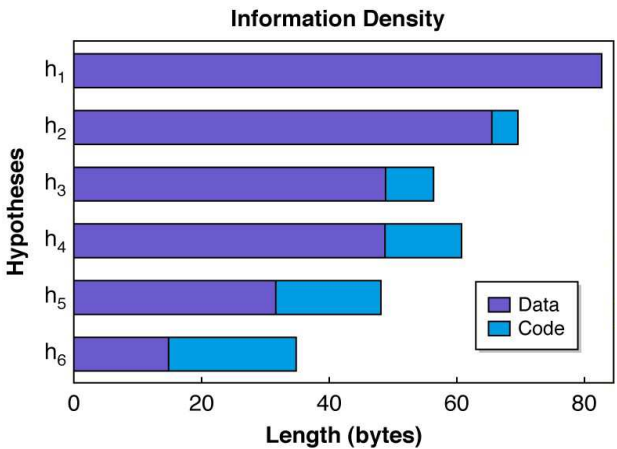,width=3.2in}}
  \caption{\label{algcont}Algorithmic Content.}
\end{figure}

\Section{Introduction}
Kolmogorov Complexity ($K(x)$) \cite{Li93a} is the optimal compression
of string $x$. This incomputable, yet fundamental property of
information has vast implications in a wide range of applications
including system management and optimization \cite{kulkANStrept,
kulkKC}, security \cite{BushKIA,EvansDISCEX01}, and
bioinformatics. Active networks \cite{BushANBook} form an ideal
environment in which to study the effects of tradeoffs in algorithmic
and static information representation because an active packet is
concerned with the efficient transport of both code and data. A
question active network application developers must answer is: ``How
can I best leverage the capabilities that active networks have to
offer?''. Because the word ``active'' in active networks refers to the
ability to dynamically move code and modify execution of components
deep within the network, this typically leads to another question:
``What is the optimal proportion of content for an active application
that should be code versus data?''. A method for obtaining the answer
to this question comes from direct application of Minimum Description
Length (MDL) \cite{Wallace:1999:MML} to an active packet. Let $D_x$ be
a binary string representing $x$. Let $H_x$ be a hypothesis or model,
in algorithmic form, that attempts to explain how $x$ is formed. Later
in this paper, we view $H_x$ as a predictor of $x$ in the analysis of
Active Virtual Network Management Prediction. For now let us focus on
developing a measure of the complexity of $x$. MDL states that the sum
of the length of the shortest encoding of a hypothesis about the model
generating the string and the length of the shortest encoding of the
string encoded by the hypothesis will estimate the Kolmogorov
Complexity of string $x$, $K(x)=K(H_x)+K(D_x|H_x)$. Note that error in
the hypothesis or model must be compensated within the encoding. A
small hypothesis with a large amount of error does not yield the
smallest encoding, nor does an excessively large hypothesis with
little or with no error. A method for determining $K(x)$ can be viewed
as separating randomness from non-randomness in $x$ by ``squeezing
out'' non-randomness, which is computable, and representing the
non-randomness algorithmically. The random part of the string,
i.e.\ the part remaining after all pattern has been removed, represents
pure randomness, unpredictability, or simply, error. Thus, the goal is
to minimize $l(H_e)+l(D_x|H_e)+l(E)$ where $l(x)$ is the length of
string $x$, $H_e$ is the estimated hypothesis used to encode the
string ($D_x$) and $E$ is the error in the hypothesis. The more
accurately the hypothesis describes string $x$ and the shorter the
hypothesis, the shorter the encoding of the string. A series of active
packets carrying the same information are measured as shown in Figure
~\ref{algcont}. Choosing an optimal proportion of code and data
minimizes the packet length. The goal is to learn how to optimize the
combination of communication and computation enabled by an active
network. Clearly, if $K(x)$ is estimated to be high for the transfer
of a piece of information, then the benefit of having code within an
active packet is minimal. On the other hand, if the complexity
estimate is low, then there is great potential benefit in including it
in algorithmic form within the active packet. When this algorithmic
information changes often and impacts low-level network devices, then
active networking provides the best framework for implementing
solutions (a specific example of separating non-randomness from
randomness, although not explicitly stated as such, can be found in
predictive mobility management as discussed in \cite{BushANBook,
liumma}. However, the optimization of code/data or
computation/communication has an additional constraint, namely
security. Complexity also plays a significant role in the analysis of
potential vulnerability within a network as discussed later in this
paper.

An active packet that has been reduced to the length of the best
estimate of the Kolmogorov Complexity of the information it transmits
will be called the minimum size active packet. When the minimum size
active packet is executed to regenerate string $x$, the $D_x|H_e$
portion of the packet predicts $x$ using static data ($E$) to correct
for inaccuracy in the estimated hypothesis. There are interesting
relationships among Kolmogorov Complexity, prediction, compression
and the Active Virtual Network Management Prediction (AVNMP) mechanism
described in \cite{BushANBook}. Details on the operation and mechanism
of operation for AVNMP can be found in papers as early as
\cite{BushPADS99}. 
Space limitations in this paper preclude a detailed
description of operation, however, an overview of the
characteristics and properties of AVNMP as well as new experimental
results are presented and the relationships among complexity,
predictability, and compressibility and information assurance are
discussed and experimentally validated throughout this paper. The next
section provides an overview of AVNMP before discussing its
relationship to Kolmogorov Complexity. After required relevant
background on AVNMP is explained, the relationship to Complexity
Theory is developed beginning from a high level overview, then driving
down into detailed relationships and experimental results.

\Section{Active Virtual Network Management Prediction Overview}

\begin{figure*}[ht]
  \centerline{\psfig{file=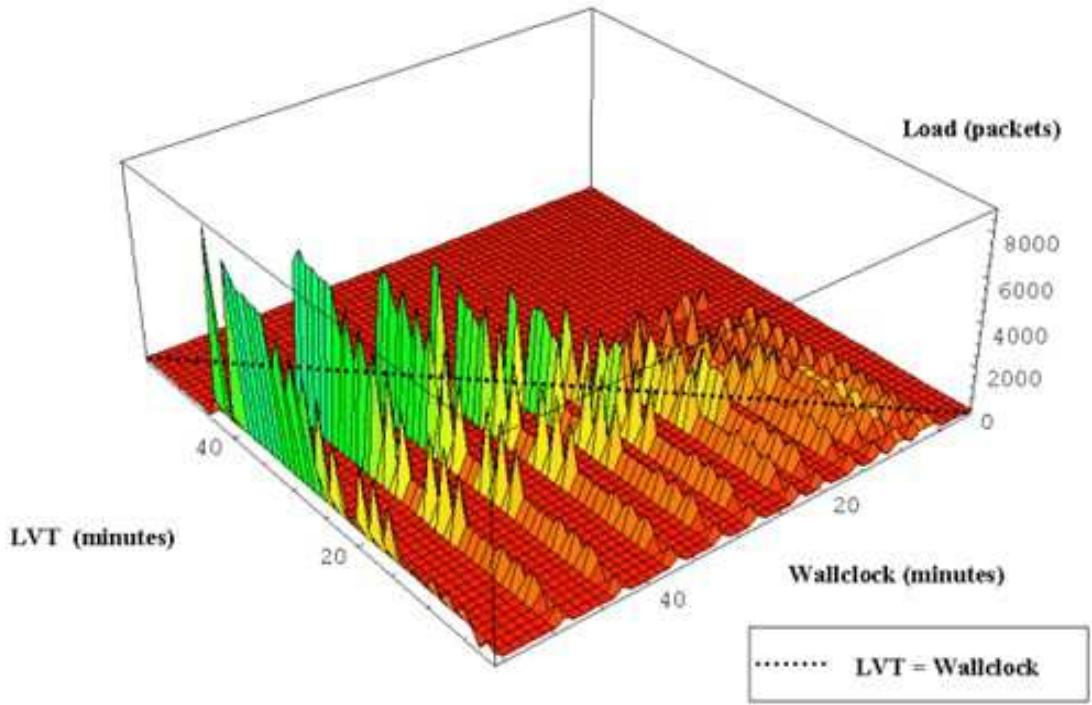,width=5.8in}}
  \caption{\label{stateq}Convergence of Prediction and Reality in the
AVNMP State Queue.}
\end{figure*}

The Active Virtual Network Management Prediction
(AVNMP)\footnote{Current project progress and experimental code is
maintained in http://www.crd.ge.com/\~{}bushsf/ftn.This research has
been funded by the Defense Advanced Research Projects Agency (DARPA)
contract F30602-01-C-0182 and managed by the Air Force Research
Laboratory (AFRL) Information Directorate.} architecture provides a
network prediction service that utilizes the capability of active
networking to inject fine-grained models into the communication
network to enhance network performance.  Active Virtual Network
Management Prediction (AVNMP) provides a network prediction service
designed to facilitate the management of large, complex, active
networks in a proactive manner. Network management includes a wide
variety of responsibilities including configuration, fault,
performance, accounting, and security management. A network management
system must be able to monitor, control, and report upon the status of
all these areas. In addition, the network management system should be
more than a tool to generate reports and help fix problems, it should
have the capability to anticipate and correct problems before they
impact network performance. AVNMP accomplishes prediction and fault
anticipation using a novel coupling of concepts from distributed
simulation and active networking. A simple example demonstrating AVNMP
results on a single node for load prediction is shown in Figure
\ref{stateq}. In today's management systems, a Management Information
Base (MIB) maintains only current state values. In AVNMP, load is
predicted into the future as real-time, called Wallclock,
advances. Thus anticipated future values are available on the node as
well as current values. In Figure \ref{stateq}, the Local Virtual Time
(LVT) (future time), runs ahead of Wallclock Time (current
time). Predicted load values are refined until Wallclock reaches the
LVT of a particular value. This capability, described in detail in
this paper, has been enabled by a new proactive network management
framework combining three key enabling technologies; namely,
distributed simulation, optimistic synchronization, and active
networks. 

The predictive capability provided by AVNMP facilitates the
development of a variety of predictive applications from mobile
wireless location management and network security to improved
QoS. AVNMP provides an ideal predictive service for mobile systems to
predict their location \cite{BushANBook}. Because mobile location
can be predicted, hand-off situations are known ahead of time and
setup for hand-off can take place prior to the hand-off event
resulting in fast hand-off and improved QoS. In the domain of network
security, AVNMP can anticipate the progress of an attack along most
likely vulnerability paths and incorporate that information into
decision-making. An attack can be propagated through the system before
it actually occurs in order to determine its impact. In collaboration
with the United States National Institute of Standards and Technology,
AVNMP has been demonstrated with CPU prediction models showing the
system's ability to detect malicious active packets. Combined with the
load prediction capability of AVNMP, Denial-of-Service attacks that
use either abnormal amounts of CPU time or large numbers of small CPU
packets can be detected and stopped \cite{NistGeDARPA}. With regards
to QoS, the load application previously discussed allows resources and
routing to be better managed by anticipating traffic in order to
optimize load distribution within the network. A few additional
selected uses for AVNMP are the ability to choose an optimum
management polling interval that minimizes overhead based upon
predicted rate of change and fault probability of the monitored data
in a managed entity, fault correction before the system is impacted
and with time available to perform dynamic optimization of repair
parts, service, and solution entities such as software agent or human
co-ordination and optimal resource allocation and planning not only
for load, but also for CPU utilization that becomes significant in
active networks. AVNMP allows ``What if...?'' scenarios to become an
integral part of the network and finally, AVNMP-enhanced components
are enabled with the ability to protect themselves by taking
proactive, evasive action, such as migrating to safe hardware before
anticipated disaster occurs.

A severe limitation of state-of-the-art network management techniques
is that they are inherently reactive. They attempt to isolate the
problem and determine solutions after the problem has
occurred. Proactive management is a necessary ingredient for managing
future networks. Part of the proactive capability is provided by
analyzing current performance and predicting future performance based
on likely future events and the network's reaction to those
events. This can be a highly dynamic, intensely computational
operation. This has prevented management software from incorporating
prediction capabilities. But distributed simulation techniques take
advantage of parallel processing of information. If the management
software can be distributed, it is possible to perform computation in
parallel and aggregate the results to minimize computation overhead at
each of the network nodes. The usefulness of optimistic techniques has
been well documented for improving the efficiency of simulations. In
optimistic logical process synchronization techniques, also known as
Time Warp \cite{BushPADS99}, causality can be relaxed in order to
trade model fidelity for speed. If the system that is being simulated
can be queried in real time, prediction accuracy can be verified and
measures taken to keep the simulation in line with actual
performance. AVNMP is implemented in an active network to provide
predictive management of an active network. AVNMP is designed to
utilize the additional processing and flexibility of an active network
to provide better management of the added complexity in processing and
bandwidth in an active network. AVNMP requires extreme network
flexibility, primarily in the ability to inject fine-grained component
models into the network. A much less flexible version of AVNMP could
be implemented in legacy systems by building dedicated network
component models directly into legacy network devices such as today's
routers. However, these models would be immobile and not easily
updated or removed, most likely requiring the network device to be
taken down when models are changed or updated. A better mechanism for
using AVNMP to manage legacy networks would be to provide an active
network overlay capable of monitoring legacy nodes. AVNMP should
reside in the active network overlay providing a predictive management
service for the legacy network. This has the added benefit of
transitioning a legacy network to an active network.

AVNMP, injected into the network as an active application, is capable
of modeling load and propagating state information in a manner that
meets the demand for accuracy at a particular active node. Greater
demand for prediction accuracy is met at the cost of AVNMP
performance, that is, the ability of AVNMP to predict farther into the
future. While this paper focuses on network traffic and load
prediction, an AVNMP application to predict CPU utilization for active
network in collaboration with National Institute of Standards and
Technology \cite{NistGeDARPA, NistGeAMS, NistGeMILCOM} has been
demonstrated. The inherently distributed nature of communication
networks and the computational power unleashed by the active
networking paradigm have been used to mutual benefit in the
development of the Active Virtual Network Management Prediction
mechanism. The active network benefits from AVNMP by continuously
receiving information about potential problems before they occur.

AVNMP benefits from the active network in many ways. The first, and most
practical way is the ease of development and deployment of this novel
prediction mechanism. This could not have been accomplished so quickly
or easily given today's closed, proprietary network device
processing. Another benefit is the fact that network packets now have
the unprecedented ability to control their own processing. Great
advantage was taken of this new capability in AVNMP. Virtual messages,
varying widely in content and processing, can adjust their predicted
values as they travel through the network. Finally, active networks
add a level of robustness that cannot be found in today's
networks. This robustness is due to the ability of AVNMP system
components, which are active packets, to easily migrate from one node
to another in the event of failure --or the prediction of failure
provided by AVNMP itself. 

\begin{figure*}[ht] 
  \centerline{\psfig{file=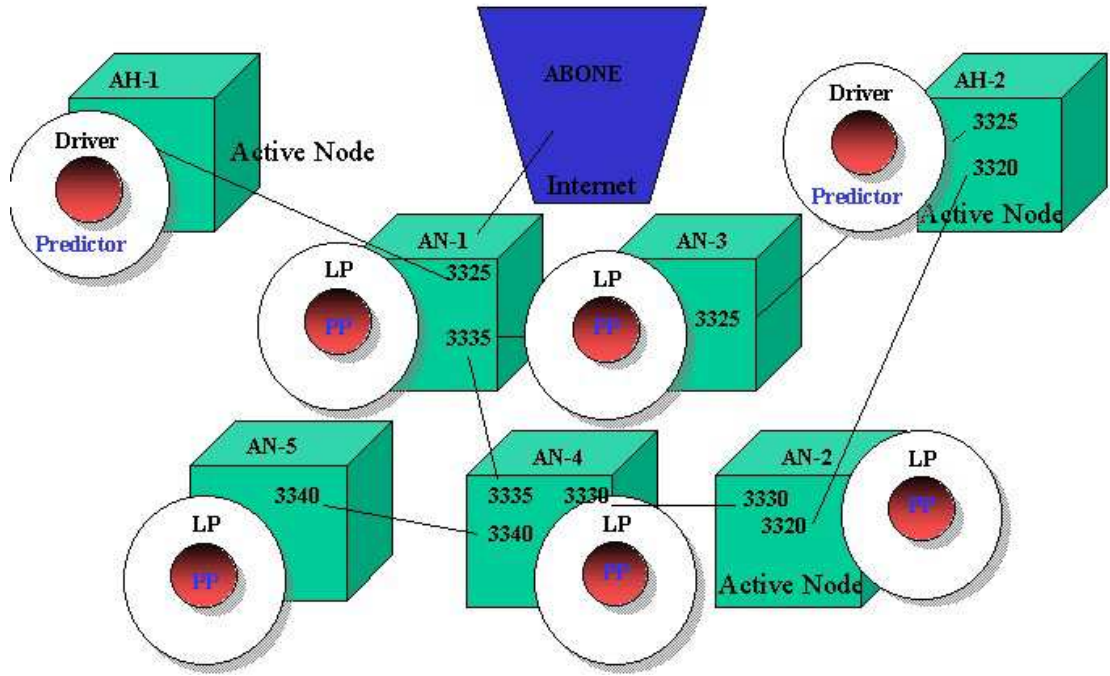,width=5.8in}}
  \caption{\label{expconfig}AVNMP Test Configuration.}
\end{figure*}

The desired characteristics of AVNMP are a large lookahead time, high
prediction accuracy, low overhead and robust operation. Each of these
characteristics is inter-related and a suitable tradeoff needs to be
determined during configuration of the system. The AVNMP experimental
validation configuration for the initial test discussed in this paper
is a feed forward network consisting of a host containing the Driving
Process and four intermediate active network nodes containing Logical
Processes as shown in Figure \ref{expconfig}. AH-1 and AH-2 are host
nodes and AN-1 through AN-5 are active network nodes. The edges
between the nodes represent links between the labeled ports on each
node. All nodes are Sun Sparcs running the Magician active network
execution environment. The AVNMP system parameters were configured as
shown in Table \ref{tab1}. In this experiment AVNMP is predicting the packet
input and output rate for each link at each node, from an application
residing on AH-1 that is transmitting an active audio packets.  

\begin{table*}[ht]
\centering
\begin{tabular}{||l|l||} \hline
{\bf Parameter} & {\bf Value} \\ \hline
Sliding Window Lookahead Length ($\Lambda$) & 200 seconds  \\ \hline
Virtual Message Generation Rate & 0.5 virtual messages/millisecond \\ \hline
Virtual Message Prediction Step Size & 20 seconds \\ \hline
Tolerance for Prediction Error ($\theta$) & 500 Messages/second \\ \hline
Virtual Real Message Ratio & 1 virtual/real message \\ \hline
Load Hypothesis ($H_e$) & Linear Extrapolation \\ 
\hline
\end{tabular}
\caption{\label{tab1}AVNMP Test Parameters.}
\end{table*}

\begin{figure*}[ht] 
  \centerline{\psfig{file=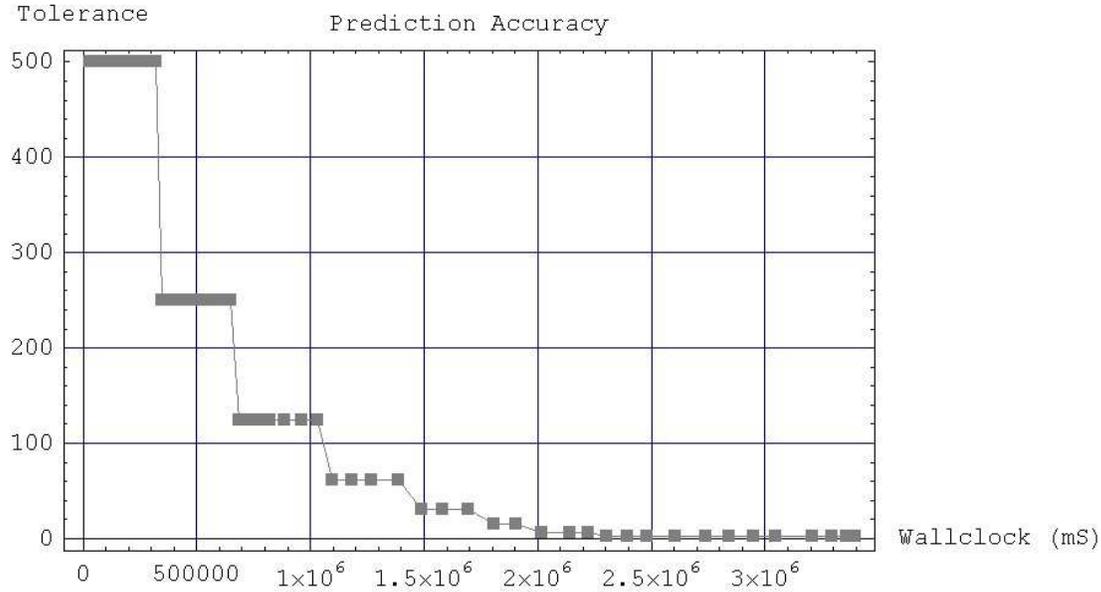,width=5.8in}}
  \caption{\label{toldec}Tolerance Setting Decreases as Wallclock
     Increases Thus Demanding Greater Accuracy.} 
\end{figure*}

\begin{figure*}[ht]
  \centerline{\psfig{file=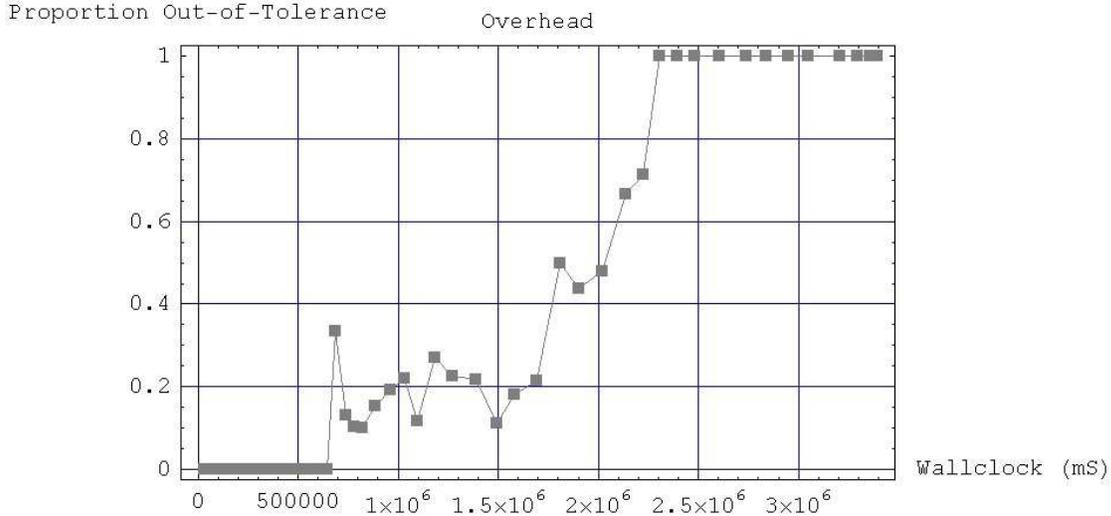,width=5.8in}}
  \caption{\label{deminc}Demand for Greater Accuracy Causes the
     Proportion of Out-of-Tolerance Messages to Increase.} 
\end{figure*}

\begin{figure*}[ht]
  \centerline{\psfig{file=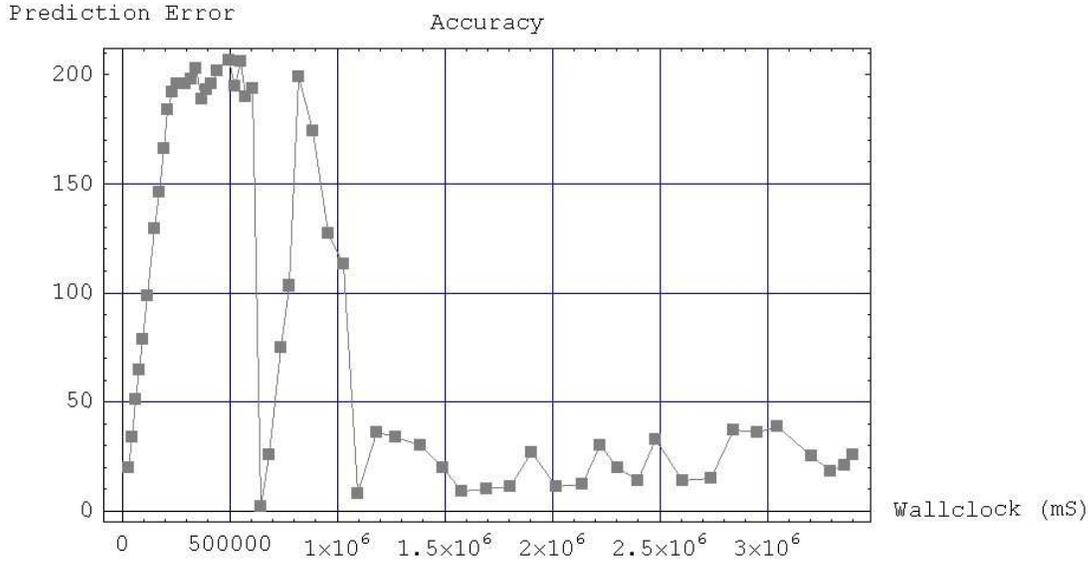,width=5.8in}}
  \caption{\label{pacc}Predictions Become More Accurate...}
\end{figure*}

\begin{figure*}[ht]
  \centerline{\psfig{file=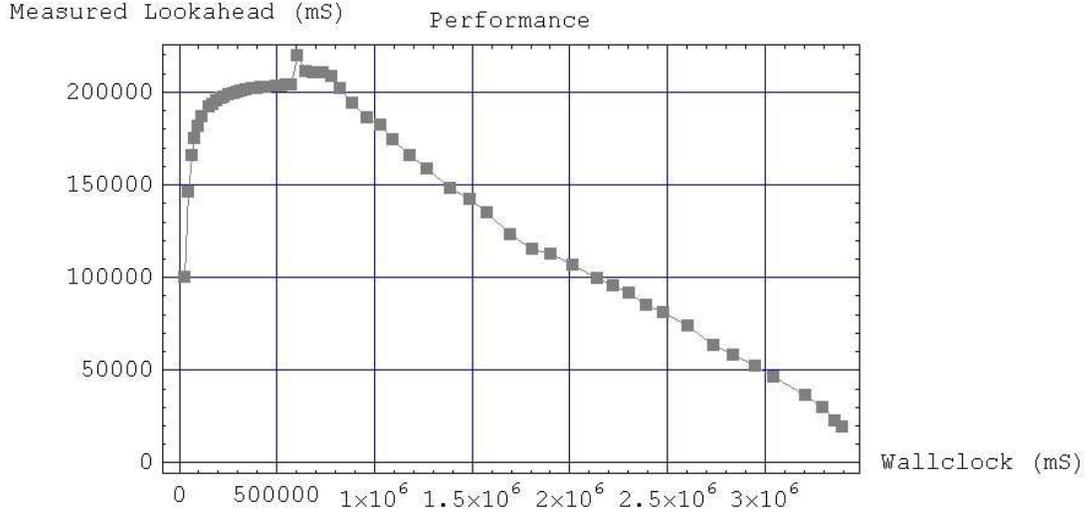,width=5.8in}}
  \caption{\label{expload}...at the Expense of Lookahead...}
\end{figure*}

\begin{figure*}[ht]
  \centerline{\psfig{file=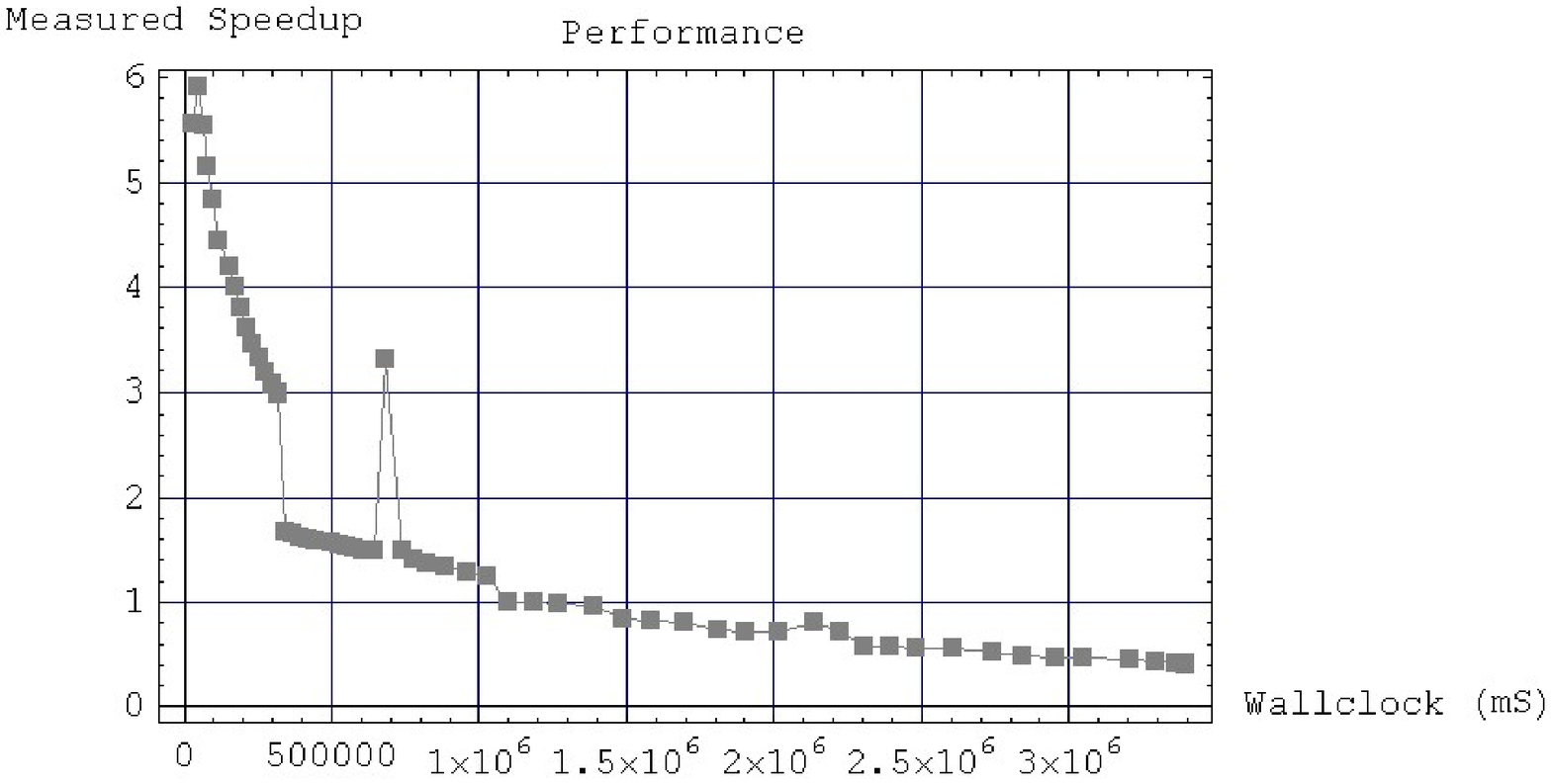,width=5.8in}}
  \caption{\label{speedup}...and Speedup.}
\end{figure*}

The State Queue plot, Figure \ref{stateq}, shows the predicted traffic
load values cached in the State Queue as a function of Local Virtual
Time (LVT) and Wallclock. As Wallclock approaches any given Local
Virtual Time, the predicted load values converge towards the actual
load. The dashed line placed diagonally across the surface highlights
where predicted time and actual time converge. The general operation
is illustrated in the next five graphs where all measurements, unless
otherwise indicated, are from node AN-4. These curves validate
intuitive trends in the operation of AVNMP. Figure \ref{toldec} shows
the reduction in tolerance versus time that is pre-programmed into
each Logical Process. The Y-axis is the tolerance that is demanded
between the predicted value and the actual value of an Simple Network
Management Protocol (SNMP) packet counter. This value is decreased
purposely in this experiment in order to create a greater demand over
time for accuracy and thus create a challenging validation of the
AVNMP system under gradually increasing stress. In Figure \ref{deminc}
the proportion of out-of-tolerance messages is shown as a function of
Wallclock. The Y-axis is the proportion of messages that arrived at a
specific node out of tolerance, that is, the actual value exceeded the
predicted value by an amount greater than the tolerance setting. As
Wallclock progresses, the tolerance is purposely reduced causing a
greater likelihood of messages exceeding the tolerance. This is done
in order to validate the performance of the system as stress, in the
form of greater demand for accuracy, is increased. Figure \ref{pacc}
shows the prediction error as a function of Wallclock. The Y-axis is
the difference in the number of packets received versus the number of
packets predicted to have been received. This graph verifies that the
system is producing more accurate predictions as the demand for
accuracy increases. However, the Y-axis of Figure \ref{expload} shows
the lookahead decreasing versus Wallclock. The expected lookahead time
is the difference between Wallclock and the Local Virtual Time at a
particular node. The demand for greater accuracy reduces the distance
into the future that the system can predict. Finally, in Figure
\ref{speedup}, speedup, the ratio of virtual time to Wallclock of the
real system, is shown as a function of Wallclock. The speedup is
reduced as the demand for accuracy is increased. As previously
mentioned, only for purposes of this experiment, the tolerance is
being reduced as Wallclock progresses, causing the accuracy to
increase while loosing performance in terms of speedup and lookahead.

\SubSection{AVNMP Overhead}

AVNMP has the potential to generate two forms of overhead, processing
overhead and bandwidth overhead. If the predicted results are within
the user specified error tolerance and the user fully utilizes the
predicted results, then overhead is at a minimum. The question of
overhead versus benefit becomes one that depends upon the perceived
utility of predictive capability and depends significantly upon the
manner and application in which it is used. It is the author's belief
that load and processing prediction are of particularly great
importance in an active network where routing is based upon not only
load, but the processing capability required by active
applications. In this section, the load prediction application example
is continued with overhead results displayed in terms of processing
time and number of packets transmitted. The expected Active Network
Encapsulation Protocol (ANEP) \cite{BushANBook} packet size measured
during this test was 1000 bytes.

\subsubsection{Task Execution Time and Message Overhead}

\begin{figure*}[ht]
  \centerline{\psfig{file=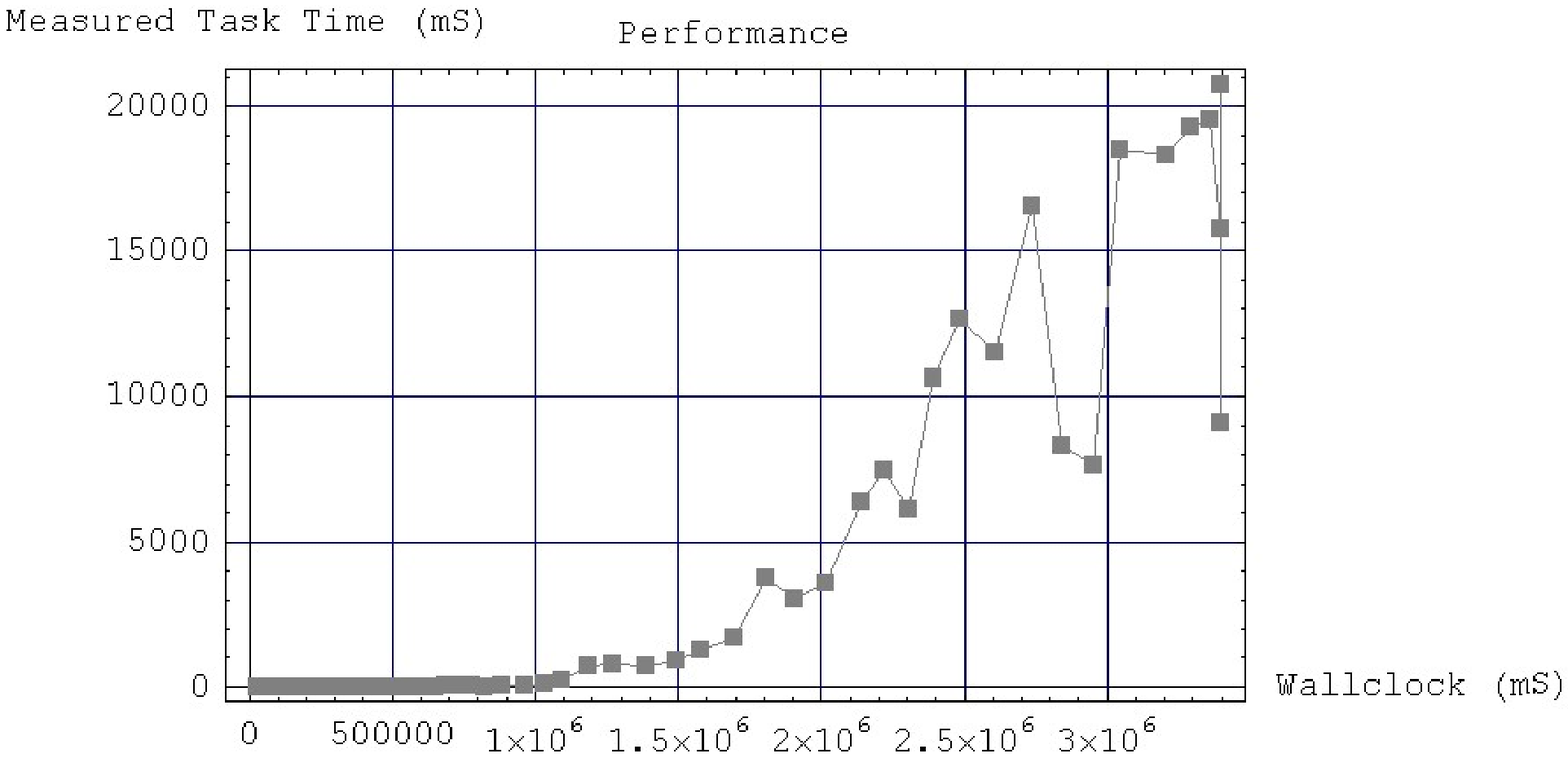,width=5.8in}}
  \caption{\label{tasktime}Expected Task Execution Time as a Function
     of Wallclock.} 
\end{figure*}

\begin{figure*}[ht]
  \centerline{\psfig{file=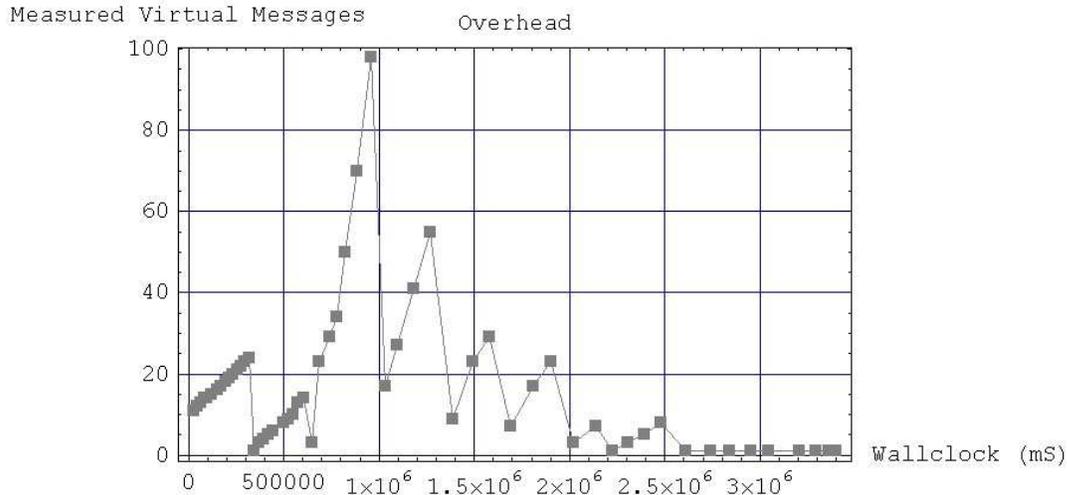,width=5.8in}}
  \caption{\label{virtmsg}Number of Virtual Messages versus Wallclock.}
\end{figure*}

\begin{figure*}[ht] 
  \centerline{\psfig{file=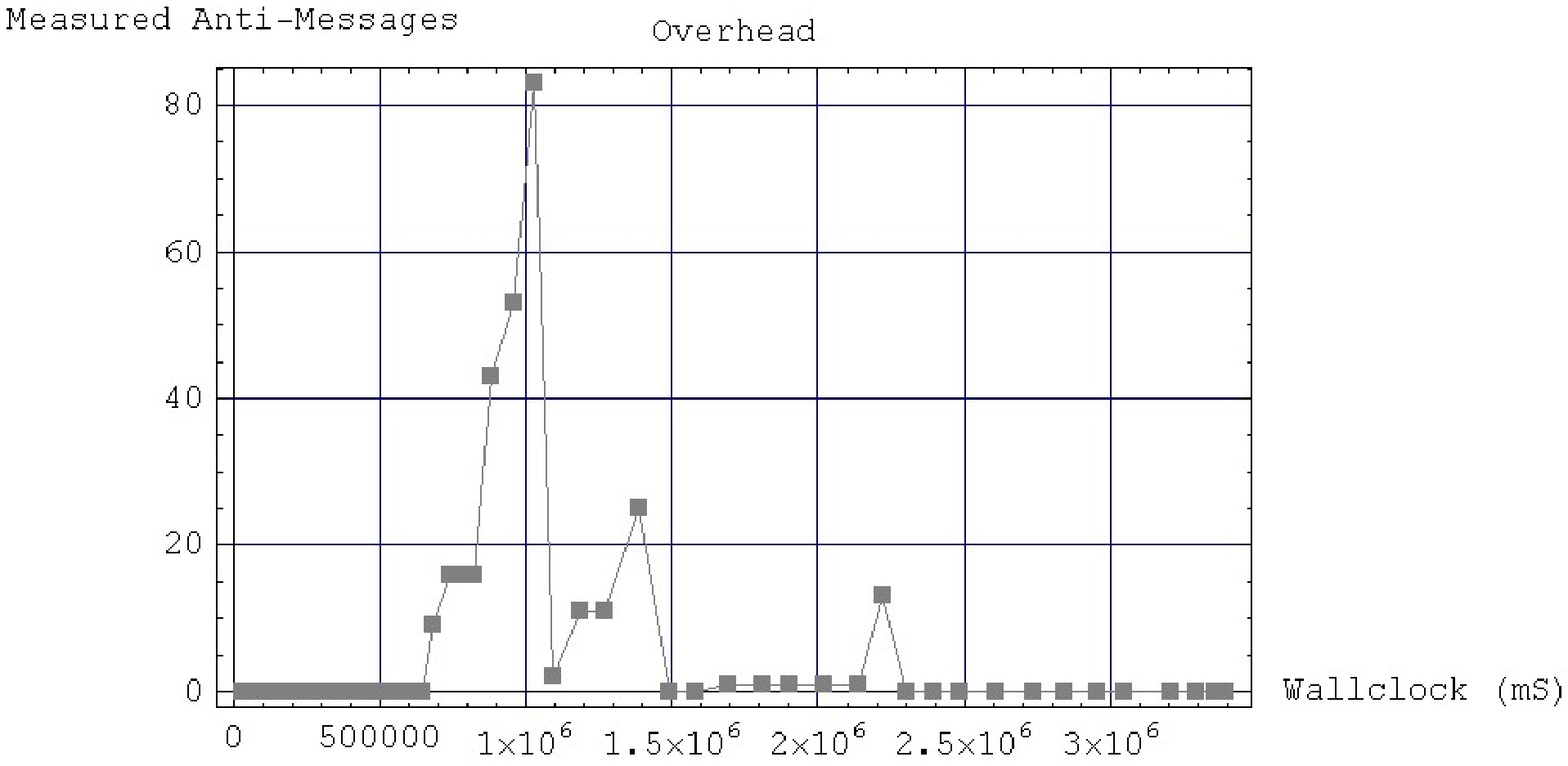,width=5.8in}}
  \caption{\label{antimsg}Number of Anti-Messages versus Wallclock.}
\end{figure*}

The task execution time is the Wallclock time the system spends
executing a non-rollback message. It was expected that task execution
time would be essentially constant; however, it increases in direct
proportion to the number of rollbacks as shown in Figure
\ref{tasktime}. This is caused by the lack of fossil collection. The
increase in the number of values in the State Queue is causing access
of the State Queue and Management Information Base (MIB) to slow in
proportion to the queue size. Figure \ref{virtmsg} displays the number
of virtual messages versus Wallclock and Figure \ref{antimsg} displays
the total number of anti-messages. This is expected to increase over
time. This value is reset every time the tolerance is tightened (every
5 minutes in this case).

\SubSection{AVNMP Robustness}
AVNMP is both an active application and an application whose purpose
is to provide predictive management of other active
applications. As a management application it must be robust in the
presence of a failing environment. So far, it has shown to provide
graceful predictive degradation in the presence of dropped packets and
broken links. AVNMP consists of two main types of
active packets: {\it AvnmpLP}, which is the Logical Process, and
{\it AvnmpPacket}, which is the virtual message. If an {\it AvnmpLP}
packet is dropped, the destination node will not have the capability
to work forward in time or forward virtual messages. Thus, AVNMP
features will not be available on the node for which {\it AvnmpLP} was
destined and the accuracy of other nodes may be reduced. If an
{\it AvnmpPacket} is dropped or unexpectedly delayed, accuracy will be
reduced because the State Queues of downstream nodes will lack a
predicted value. However, AVNMP will continue to operate with degraded
performance. In the next section the role of complexity in
understanding prediction is discussed. Ideally, of course, AVNMP
should have predicted the error condition and taken action to mitigate
it. However, control mechanisms have not yet been implemented.

\begin{figure}[ht] 
  \centerline{\psfig{file=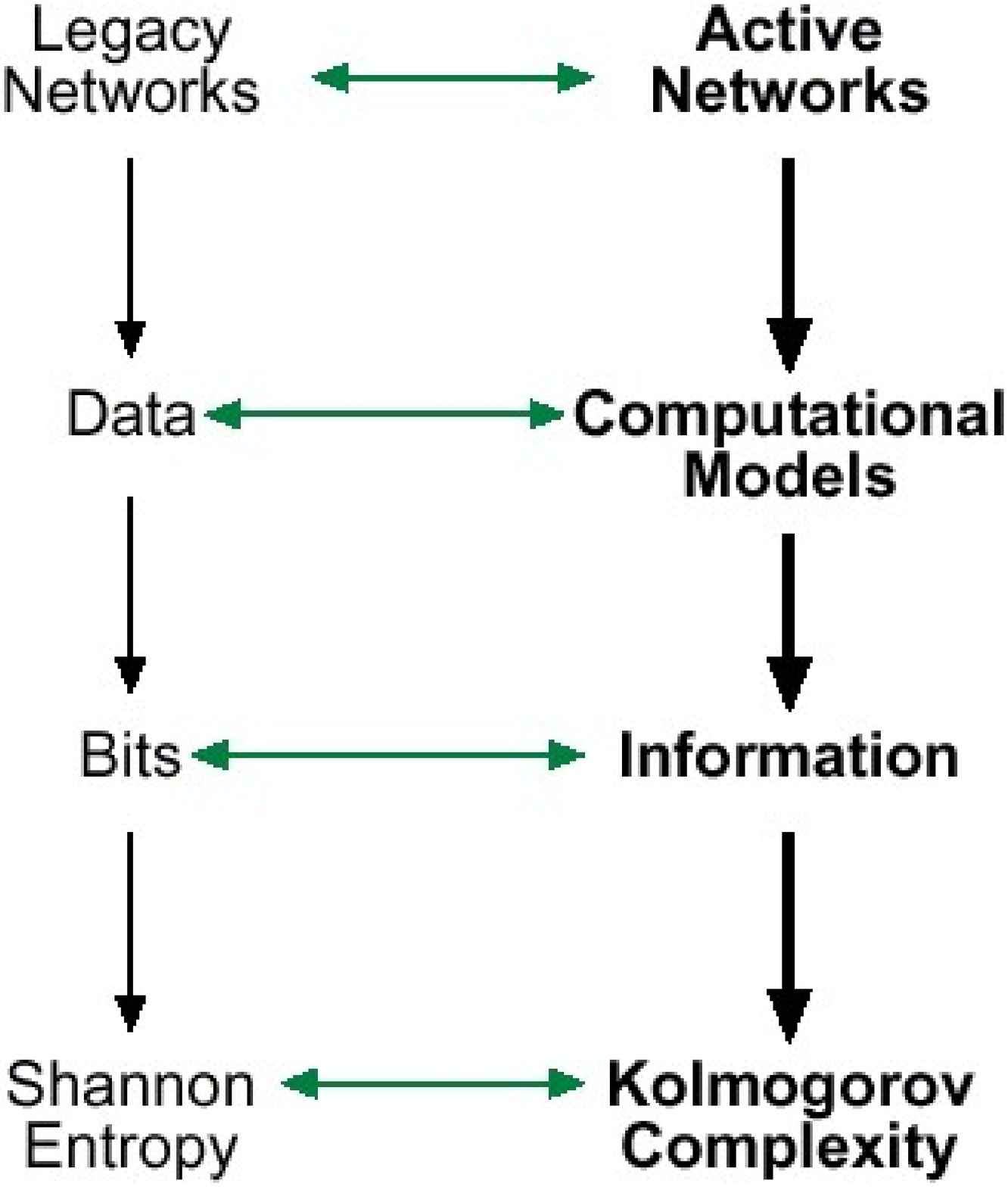,width=3.2in}}
  \caption{\label{actleg}Active Networks and Legacy Networks as
     Viewed by AVNMP.} 
\end{figure}

\SubSection{Networking Viewed Through the Lens of Complexity}
AVNMP can provide early warning of potential problems; however, the
identification of a solution and marshaling of automated solution
entities within an active network has not yet been fully
addressed. This project has begun to lay the groundwork for such
automated composition of management solutions within an active network
\cite{BushANBook}. This direction is being carried forward by
exploration of a relatively unexplored area -understanding the
benefits of active networking, Algorithmic Information Theory, and its
close companion, Complexity Theory. To our knowledge, this work is the
first to propose and begin investigation into the newly available
processing power of active networking through the concept of
Complexity and Algorithmic Information (``Streptichrons'') as shown in
Figure \ref{actleg}. Legacy networks, which are today's passive
networks, have been designed to optimize transmission of passive data
using bit compression based upon the underlying notion of Shannon
Entropy. AVNMP has shown that active networks allow for the
possibility of executable models and that the corresponding
information packets might be best studied with Kolmogorov Complexity
as the underlying theory. It is serendipitous that Complexity Theory
has been receiving more attention lately and is making significant
theoretical progress at the same time that research into active
networking is taking place. Active networks provide a new paradigm and
enhanced capabilities, which, when combined with ideas from
Algorithmic Information Theory \cite{Li93a}, might lead to superior,
innovative solutions to problems of network management. One possible
approach proposes to combine Kolmogorov Complexity with the science of
Algorithmic Information Theory (sometimes called Complexity Theory) to
build self-managed networks that draw on fundamental properties of
information to identify, analyze, and correct faults, as well as
security vulnerabilities, in a distributed information system
\cite{kulkANStrept,kulkKC}. Specifically, we suspect that complexity
measures can be used to detect and analyze problems in a network, and
to facilitate techniques to remedy network faults. We also envision
that Kolmogorov Complexity can be applied directly to improve the
performance of AVNMP. In general, complexity is not computable;
however, the bounds on complexity tighten continuously as fundamental
research in Kolmogorov Complexity progresses.

One potential drawback to AVNMP, gently pointed out earlier in this
paper, is the fact that AVNMP itself consumes resources in an effort
to predict resource usage in a network. Resource consumption by AVNMP
is tied directly to accuracy: higher accuracy costs more in terms of
bandwidth utilization, associated with simulation rollbacks and the
concomitant transmission of anti-messages. Despite this relationship,
potential exists to nearly reach the theoretical minimum amount of
bandwidth to achieve maximal model accuracy. This possibility arises
because AVNMP consists of many small, distributed models (each a
description of a theory) that work together in an optimistic,
distributed manner via message passing (data). Each AVNMP model can be
transferred, using an active network, as a Streptichron
\cite{BushANBook}, which is any message that contains an executable
model in addition to data for prediction. Using Streptichrons, the
optimal mix of data and model can be transmitted to implement
MDL. Achieving maximal model accuracy at minimal bandwidth provides
the best AVNMP accuracy at the least cost in AVNMP resource
consumption.

Other possibilities exist to exploit Kolmogorov Complexity to improve
AVNMP performance. For example, one can apply the MDL technique to the
rollback frequency of all the AVNMP enhanced nodes in a network. A low
rollback complexity (which suggests a high compressibility in the
observed data) would indicate patterns in the rollback behavior that
could be corrected relatively easily by tuning AVNMP parameters. High
complexity (low compressibility) would indicate the lack of any
computable patterns, and would suggest that little performance
improvement could be achieved by simply tuning parameters. Thus, we
hypothesize that our tuning gradient should be guided toward regions
of high complexity, which suggests that we can tune parameters to
improve the rollback frequency. The next section focuses upon
experimental results relating prediction to complexity gathered from
the operation of the AVNMP system. 

\begin{figure*}[ht] 
  \centerline{\psfig{file=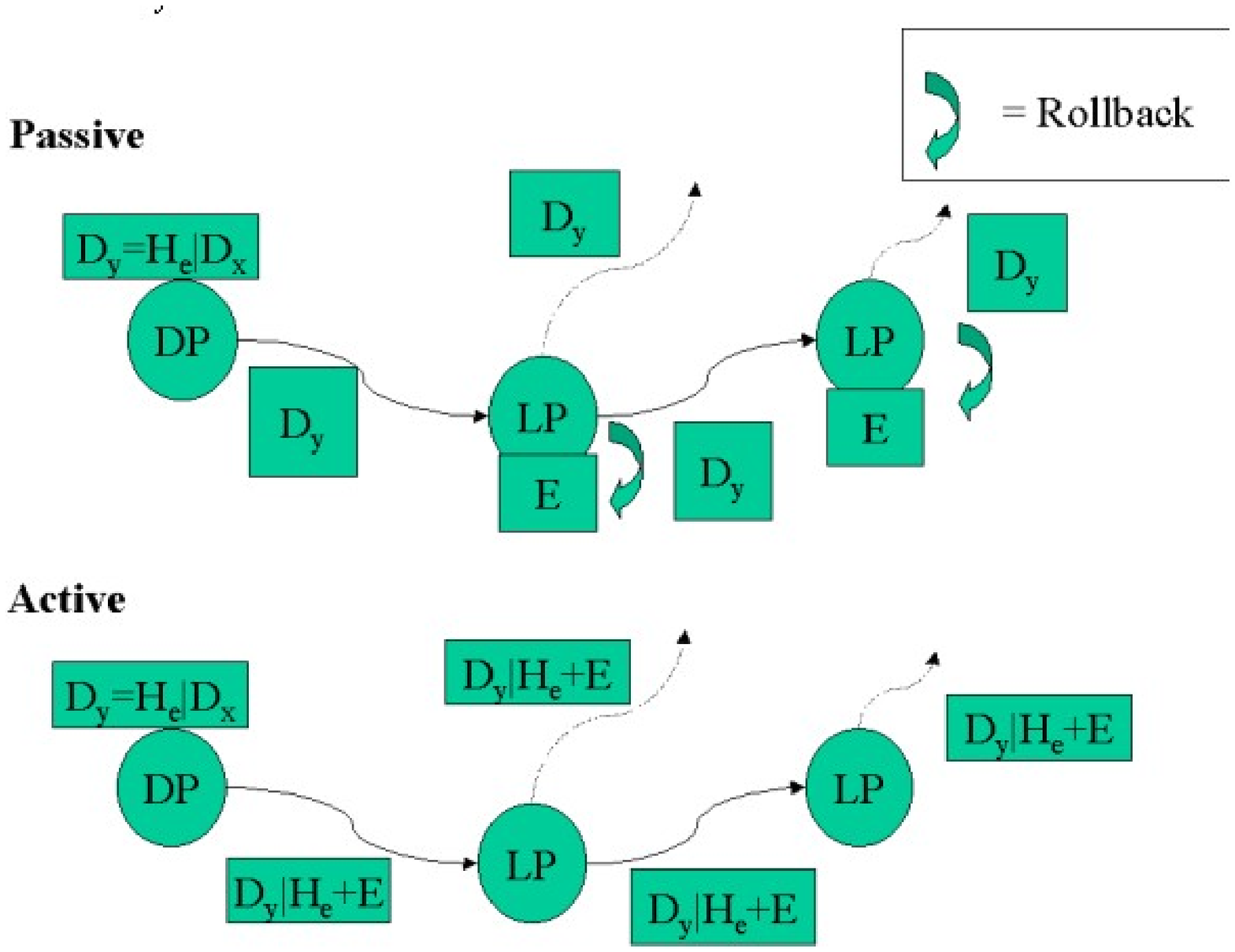,width=5.8in}}
  \caption{\label{actpassive}Active versus Passive Form of AVNMP.}
\end{figure*}

\SubSection{AVNMP and Kolmogorov Complexity}
In AVNMP, information that impacts the network is transmitted based
upon prediction at a low level within the network. Thus, AVNMP allows
experimentation in defining the boundaries within which active
networking is beneficial. In Figure \ref{actpassive} an active and
passive form of AVNMP is represented. The passive case is represented
in the upper portion of the figure. In the passive case, actual data
($D_x$) is observed at the Driving Process.  Note that in the AVNMP
architecture, Driving Processes exist at the edge of the system. They
monitor external forces acting upon the system, such as load, and
generate virtual messages, which are a short-term local prediction
about a specific property such as input load, that are injected into
the AVNMP system. The Driving Process has a hypothesis that has been
formed about the data; predicted data ($D_y$) is generated in the form
of static virtual messages. The term static indicates that information
content within the message contains no executable code. The virtual
messages are propagated through the network driving the the system
ahead of Wallclock. When error in the hypothesis exceeds a preset
threshold, AVNMP causes rollbacks to occur in order to adjust for the
inaccuracy. In the lower portion of Figure \ref{actpassive}, the
hypothesis is included within each packet and is used to encode $D_y$
within the code portion of the active packet.

\begin{figure}[ht] 
  \centerline{\psfig{file=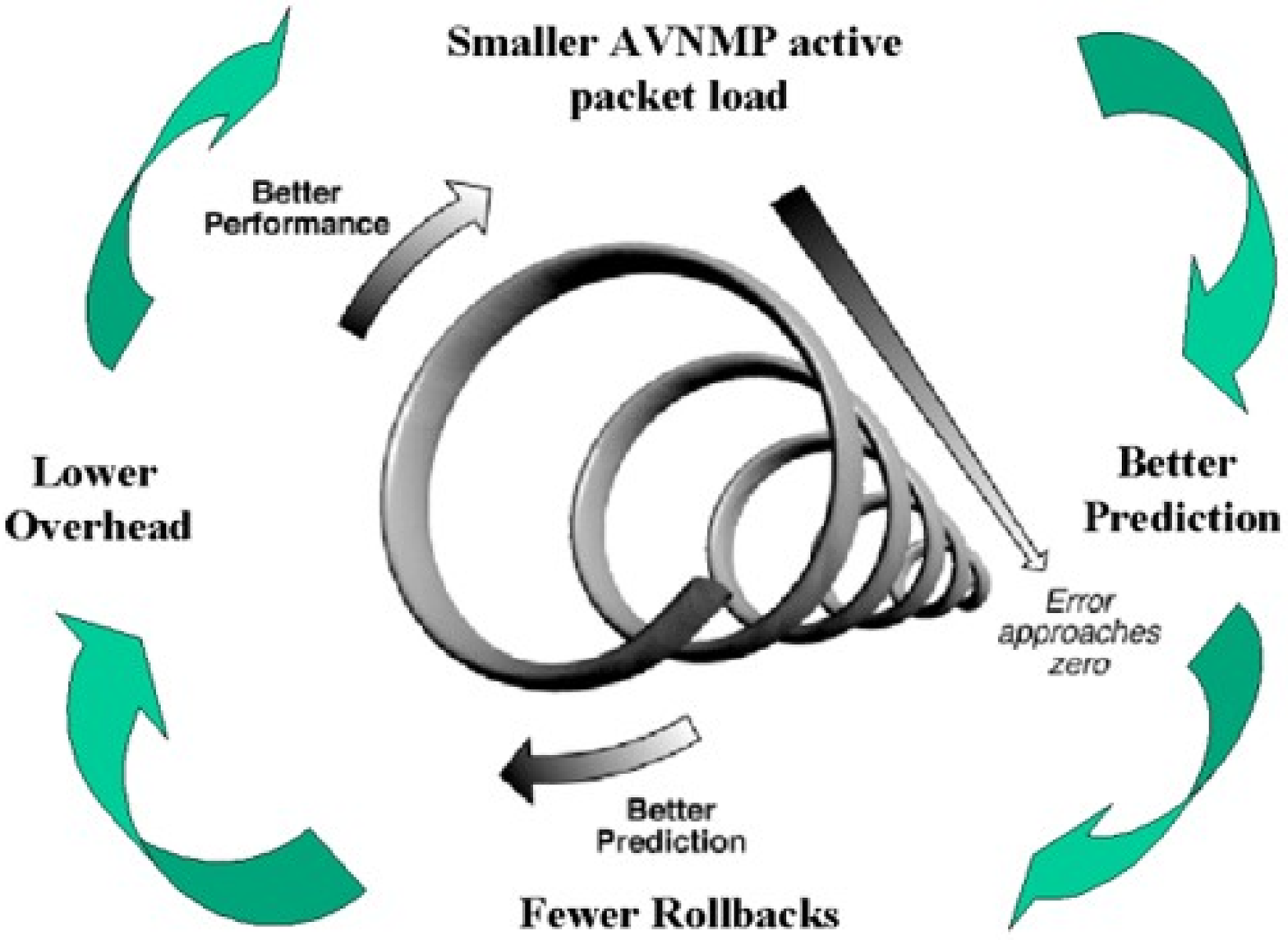,width=3.2in}}
  \caption{\label{selfimprove}Better Prediction Implies Smaller
     Packets Implies Better AVNMP Performance Implies Better
     Prediction.} 
\end{figure}

What is the relationship between the estimated operating hypothesis
($H_e$) that can encode an AVNMP packet and $H_e$ as the predictor in
the Driving Processes? First, they are the same hypothesis. Second, it
has been shown \cite{Li93a} that the shorter the packet, the better
the predictor. Conversely, the worse the prediction, the longer the
$E$ value, where $E$ can be considered any of the following equivalent
names: error, complexity, or randomness within the AVNMP packet
encoding. Can Active Virtual Network Management Prediction benefit
from the fact that the smallest algorithmic form is also the most
likely predictor of a sequence? This can come about because Driving
Processes and Streptichrons (active virtual messages anticipating
events in the future) benefit by being both small and accurate as
shown in Figure \ref{selfimprove}. The objective is to increase the
rate of convergence of the predictions held within the State Queue to
converge to the actual value that will occur in the future, and to
converge to that value before it actually exists. Actual and predicted
values within a particular instance of a State Queue were shown in
Figure \ref{stateq}. Let us examine AVNMP results in light of
complexity in more detail in the next section.

\SubSection{Load Prediction And Complexity In Active Virtual Network
     Management Prediction} 

With regard to active packets and information theory, passive data is
simple Shannon compressed data, and active packets are a combination
of data and program code whose efficiency can be estimated by means of
Kolmogorov Complexity. The active network Kolmogorov Complexity
estimator is currently implemented as a quick and simple compression
estimation method. It returns an estimate of the smallest compressed
size of a string. It is based upon computing the entropy of the weight
of ones in a string. Specifically it is defined in Equation \ref{eq:1}
where $x\#1$ is the number of 1 bits and $x\#0$ is the number of 0 bits
in the string whose complexity is to be determined. Entropy is defined
in Equation \ref{eq:2}. See \cite{EvansOSCR} for other measures of
empirical entropy and their relationship to Kolmogorov Complexity. The
expected complexity is asymptotically related to entropy as shown in
Equation \ref{eq:3}.

\begin{eqnarray}
\label{eq:1}\hat{K}(x)=l(x) H({{x\#1}\over{x\#1+x\#0}})+\log_2(l(x))\\
\label{eq:2}H(p)=-p\log_2p-(1.0-p)\log_2(1.0-p)\\
\label{eq:3}H(X)=\sum_{l(x)=n}P(X=x)K(x)
\end{eqnarray}

\begin{figure}[ht] 
  \centerline{\psfig{file=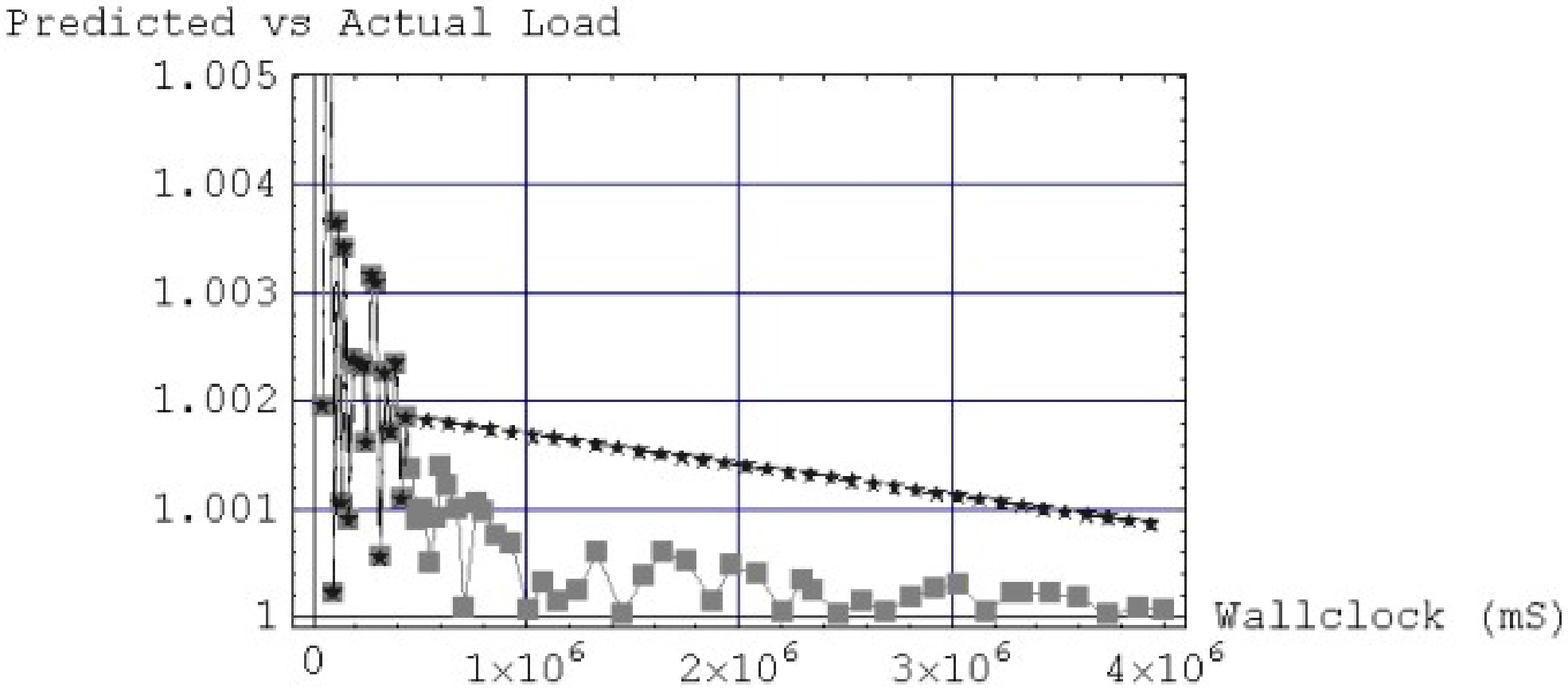,width=3.2in}}
  \caption{\label{loadpred}Prediction Hypothesis ($H_e$) Compared
     with Actual Load in AVNMP Test.}
\end{figure}

Load prediction data sampled from execution of AVNMP is analyzed
relative to several hypotheses. The goal is to use a simple example to
demonstrate the relationship among accuracy of hypotheses, complexity,
and compression. The initial hypothesis ($H_e$) (regardless of naivete
in choice of hypothesis) is that the data can be characterized by a
simple linear extrapolation based upon the last sampled load
values. This is shown in Figure \ref{loadpred} where the gray boxes
are actual load samples and the black stars are predicted load
samples. Note that the predicted load is based upon a short history
shown in the graph as the initial match between predicted and actual
load.

\begin{figure}[ht] 
  \centerline{\psfig{file=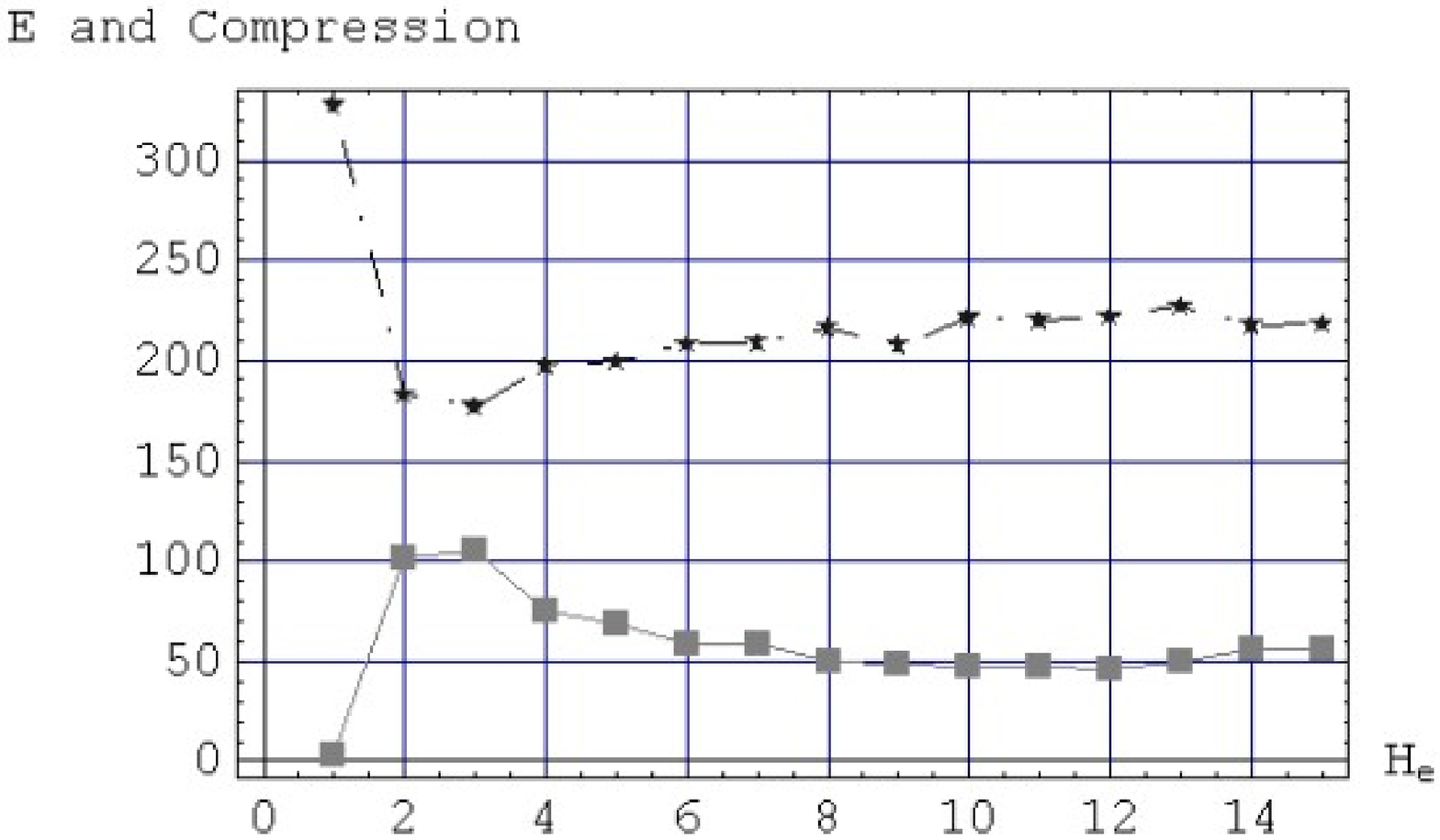,width=3.2in}}
  \caption{\label{AVNMPhyp}Prediction Error and Complexity Estimate
     over a Range of AVNMP Hypotheses for Load Prediction.}
\end{figure}

Various enhancements are added to the initial hypothesis to create new
hypotheses for our test. In this specific case, a running average was
used to smooth the data before the extrapolation. The size of the
running average defines a hypothesis. Each enhancement is considered a
new hypothesis ($H_e$) in this experiment. In Figure \ref{AVNMPhyp},
for each $H_e$ the sum of the error in predictions is graphed as the
gray boxes in the lower portion of the graph. The compressed size of
the corresponding error is plotted as the black stars in the upper
portion of the figure. Clearly a better hypothesis concerning the
origination of the data results in better prediction and greater
compression, while poor hypotheses result in inaccurate prediction and
reduced compression. This provides a concrete demonstration of the
relation between complexity and prediction accuracy.

\begin{figure*}[ht]
  \centerline{\psfig{file=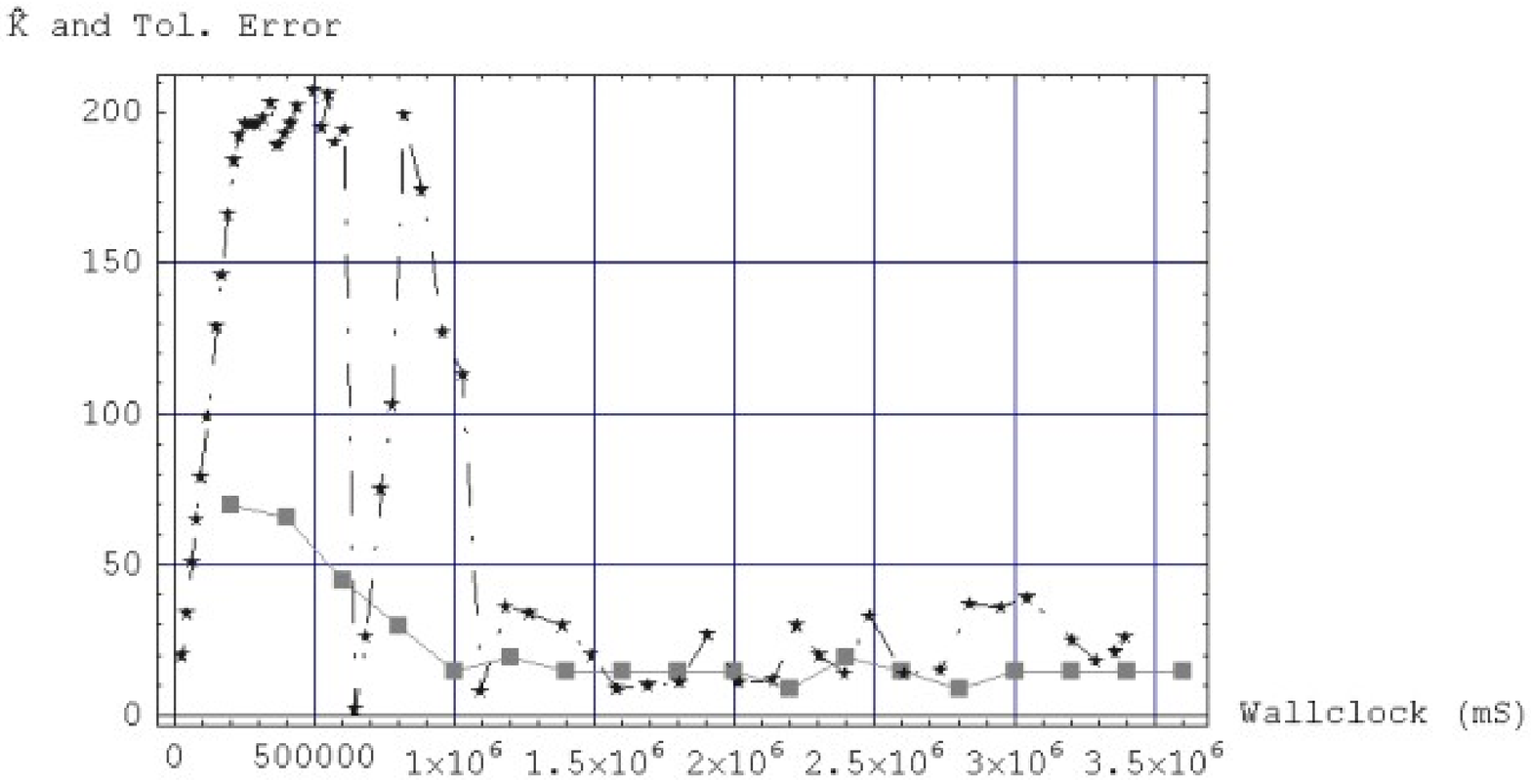,width=5.8in}}
  \caption{\label{KvsE}Estimated Complexity and Error within AVNMP.}
\end{figure*}

A key contribution presented in this paper is the hypothesis and
supporting experimental validation that greater complexity results in
greater prediction error, and thus greater likelihood of AVNMP
rollback. Load prediction error from AN-1 (see the experimental
configuration shown in Figure \ref{expconfig}) within the network is
compared with the estimated complexity of the actual load. In Figure
\ref{KvsE} the load prediction error is plotted with the estimated
complexity versus Wallclock where values are taken over intervals of
the same length as the Sliding Lookahead Window shown in Table
\ref{tab1}. Larger error, and thus more likely rollback, occurs during
periods of relatively high complexity, while complexity is low during
periods of low prediction error.

As predictions become more inaccurate in AVNMP, virtual messages
should slow down, rather than burden the system with potential
rollbacks. Poorly predicted messages will naturally be larger in their
minimum size, which slows down their rate of propagation in proportion
to their inaccuracy. 

Another issue concerns a mechanism for feedback to the Driving Process
in order to improve $H_e$. Such a feedback mechanism can be based upon
input from the complexity estimate, or minimum encoded packet size, of
virtual messages. The hypothesis is adjusted in a manner that drives
the system towards minimizing encoded virtual message size. 

\Section{Complexity and Assurance}
Complexity is useful not only for management prediction and active
packet length optimization, but also for security. The vulnerability
analysis technique presented in this section takes into account the
innovation of an attacker. A metric for innovation is not new; 700
years ago William of Occam suggested a technique \cite{Kircher}. The
salient point of Occam's Razor and complexity-based vulnerability
analysis is that the better one understands a phenomenon, the more
concisely the phenomenon can be described. This is the essence of the
goal of science: to develop theories that require a minimal amount of
information to be fully described. Ideally, all the knowledge required
to describe a phenomenon can be algorithmically contained in formulae,
and formulae that are larger than necessary lack of a full
understanding of the phenomenon. Consider an attacker as a scientist
trying to learn more about his environment, that is, the target
system. Parasitic computing \cite{Parasitic} is a literal example of a
scientist studying the operation of a communication network and
utilizing it to his advantage in an unintended manner. In Parasitic
computing, checksums, additional overhead supposedly designed to
insure the integrity of the information, are turned against the system
and used to the attacker's advantage. In fact, because information
assurance safeguard developers do not yet have a comprehensive
conceptual framework in which to evaluate the effectiveness of their
safeguards individually or as composites of safeguards, such scenes
are unfortunately all too common. Safeguard designers must be able to
capture and quantify the mechanism by which an attacker as scientist
generates hypotheses and theorems about a system under
attack. Theorems are attempts to increase understanding of a system by
assigning a cause to an event, rather than assuming all events are
randomly generated. If theorem $x$, described in bits, is of length
$l(x)$, then a theorem of length $l(m)$, where $l(m)$ is much less
than $l(x)$, is not only much more compact, but also $2^{l(x)-l(m)}$
times more likely to be the actual cause than pure chance
\cite{Kircher}. Thus, the more compactly a theorem can be stated, the
more likely the theorem is to be correct. A measure of this
compactness is described and utilized in more detail later in this
paper.

Imagine a vulnerability identification process that consists of the
following: waiting for an information system to be attacked, then,
assuming it survives and one can detect the attack, analyzing the
attack, and if the information system is still not compromised, adding
this information to one's knowledge base. This technique would be
unacceptable to most people, but it is essentially the technique used
today. Information assurance, and vulnerability analysis in
particular, are hard problems primarily because they involve the
application of the scientific method by a defender to determine a
means of evaluating and thwarting the scientific method applied by an
attacker. This self-reference of scientific methods would seem to
imply a non-halting cycle of hypothesis and experimental validation
being applied by both offensive and defensive entities, each affecting
the operation of the other. Information assurance depends upon the
ability to discover the relationships governing this cycle and then
quantifying and measuring the progress made by both an attacker and
defender. The salient factor controlling the paths taken by attacker
and defender are governed by the complexity of the system. Whether
such properties are measurable and how they will behave in a complex
system is a topic of open debate. However, a metric and framework are
required for quantifying information assurance in such an environment
of escalating knowledge and innovation. Progress in vulnerability
analysis and information assurance research cannot proceed without
fundamental metrics. The metrics should identify and quantify
fundamental characteristics of information in order to guarantee
assurance.

\begin{figure*}[ht] 
  \centerline{\psfig{file=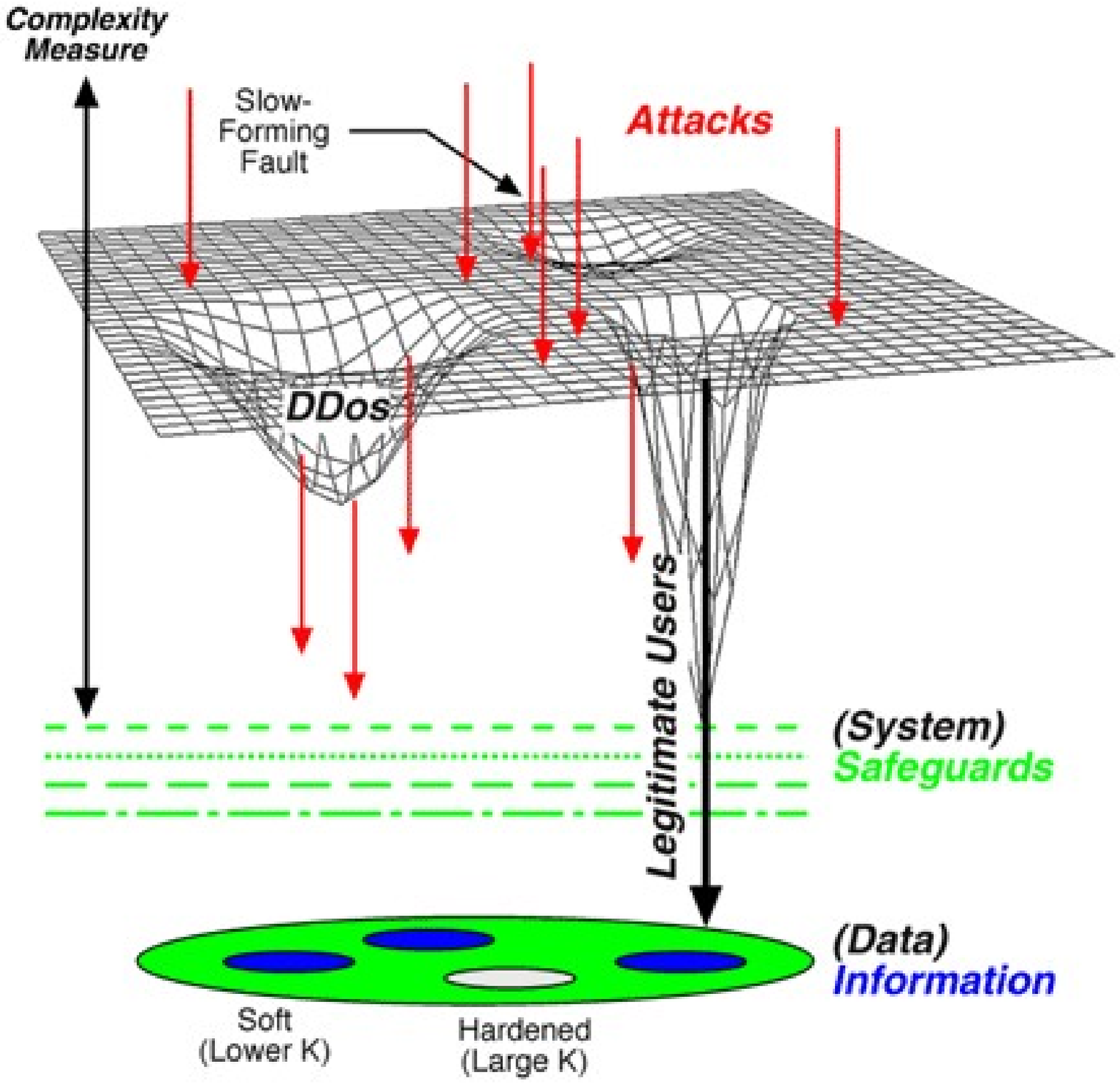,width=5.8in}}
  \caption{\label{IAoverview}Conceptual Illustration of a
Vulnerability and Attack Detection Complexity Grid.}
\end{figure*}

A fundamental definition of vulnerability analysis is formulated in
this paper based upon attacker and defender as reasoning entities,
capable of innovation. Truly innovative implementations of attack and
defense lead to the evolution of complexity in an information
system. Understanding the evolution of complexity in a system enables
a better understanding of where to measure and how to quantify
vulnerability and leads towards a calculus of information
complexity. The design and implementation of a complexity-based
vulnerability analysis technique is under development as a tool for
automated measurement of information assurance. The motivation for
complexity-based vulnerability analysis comes from the fact that
complexity is a fundamental property of information and can be
ubiquitously applied. The presentation and analysis of the Kolmogorov
Complexity-based vulnerability analysis framework must accomplish
several goals. As initially stated, the vulnerability analysis
technique must demonstrate the ability to account for the innovation
of an attacker. The technique should be based upon fundamental
properties of information, rather than suffer from the combinatorial
explosion that occurs when explicitly examining all possible events
generated by specific systems. The vulnerability results should make
intuitive sense; vulnerability is reduced by increasing the apparent
complexity of access to information from potential attackers while
increasing vulnerability for less complicated, or in some sense
shortest paths of access to information. A topological view of
vulnerability can be demonstrated. This is demonstrated by means of a
Kolmogorov Complexity Map (K-Map) in which low complexity paths, which
are likely to be easy for an attacker to follow, are identified. The
concept of a K-Map, or complexity grid, is shown in Figure
\ref{IAoverview} and the K-Map for a specific example is derived later
in this paper and shown in Figure \ref{Kcontour}. Figure
\ref{IAoverview} may itself appear quite complex upon first glance;
however, focus upon individual parts of the figure in a logical
progression. Begin with the information to be protected that lies at
the bottom of Figure \ref{IAoverview}. Attacks are illustrated as the
thin downward-pointing arrows attempting to penetrate the system in
order to manipulate the information. Numerous safeguards, supposedly
designed to protect the information, each designed to mitigate
particular types of attack, are shown as barriers with various levels
of porosity inserted across the middle of the figure. The overall
complexity of the system is illustrated by the surface contour located
above the information and safeguards and is comprised of the
complexity of several entities, namely: the information itself, the
complexity of the system in which the information resides and the
complexity of the safeguards. Innovative attacks will be more likely
to successfully penetrate areas of low complexity, easier to
comprehend components of the system. In addition, specific types of
attacks, such as Distributed Denial of Service (DDoS) will appear as
warps in the complexity grid. This is due the inherent system
correlation in DDoS attack-streams. The vulnerability analysis
technique should be applicable in a highly dynamic and amorphous
information environment; an active network environment is chosen
because information can be transmitted through an active network while
its proportion of algorithmic content varies. In other words, static
data or executable code or various combinations of both can represent
information; both forms of information should have high assurance. The
assurance of their interaction at a low level within an active network
presents a nice challenge.

Kolmogorov Complexity provides a measure of complexity that can be
utilized for vulnerability analysis. Observe an input sequence at the
bit-level and concatenate with an output sequence at the
bit-level. This input/output concatenation is observed for either the
entire system or for components of the system. Low complexity
input/output observations quantify the ease of understanding by a
potential attacker. Previous work has demonstrated the use of
Kolmogorov Complexity for Distributed Denial of Service (DDoS) attack
detection \cite{AmitDDoS}. Definition 1 explicitly states the means of
measuring the complexity of a system component, or protocol
interaction, to a potential attacker.

\begin{description}
\item[Definition 1:Vulnerability Metric] Vulnerability is inversely
  proportional to $K(x[opstart:opend]) /l(x[opstart:opend])$ where
  $opstart$ is the bit at which an operation to be discovered within
  an information system begins, and $opend$ is the last bit in an
  attacker's observation.
\end{description}

\begin{figure}[ht] 
  \centerline{\psfig{file=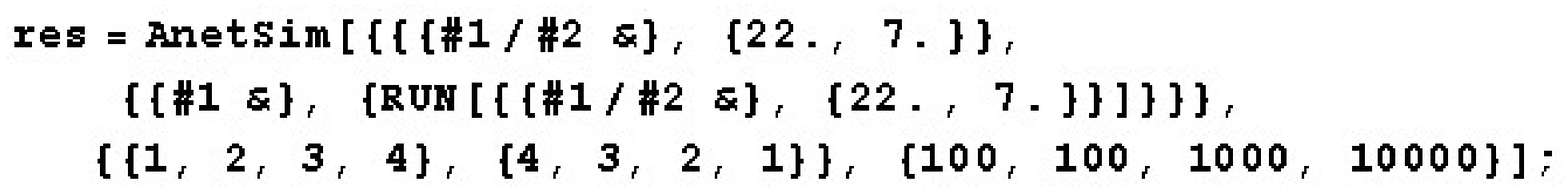,width=3.2in}}
  \caption{\label{algview}Algorithmic View of Active Packets.}
\end{figure}

\begin{figure*}[ht]
  \centerline{\psfig{file=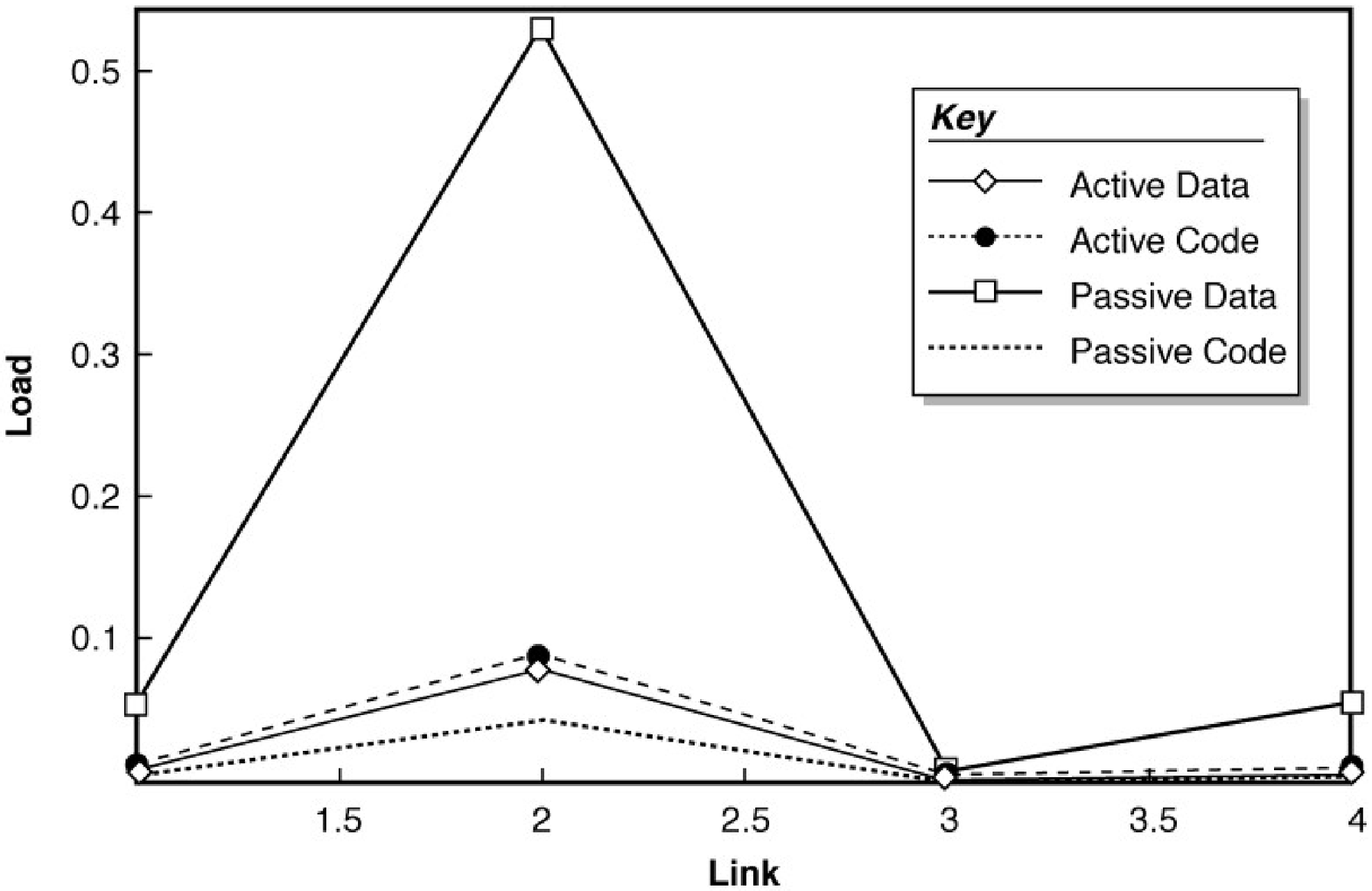,width=5.8in}}
  \caption{\label{Kloadgraph}Algorithmic versus Static Active Network
Information Load.}
\end{figure*}

In the remainder of the paper, excerpts from a Mathematica Notebook
are included. The excerpts contain code using common mathematical and
programming constructs and therefore should be intuitively obvious
without requiring knowledge specific to Mathematica; any Mathematica
specific details are explained in the text. As a specific example of
the algorithmic capabilities of active networks, consider the
transmission of an estimate of $\pi$. One could choose to send $\pi$
as an extremely large number of digits; in contrast, one could send a
smaller algorithm capable of generating $\pi$ to an arbitrary number
of digits. Consider an illustration of this concept in more
detail. The Mathematica code, $\{\{\#1/\#2\&\}\},\{22.,7.\}\}$,
represents an unnamed function that divides the first argument by the
second argument; the function implements $22/7$. Consider that the
function ($\{\{\#1/\#2\&\}\}$) and the data ($\{22.,7.\}$) remain
unevaluated and are transmitted together. This represents an active
packet; it contains part code and part data. The RUN function
evaluates the function and returns the result; the result in this case
is static data, a legacy data packet. Mathematica code that analyzes
the characteristics of algorithmic and passive information
transmission is shown in Figure \ref{algview}. The active packet is
defined as $\{\{\#1/\#2\&\}\},\{22.,7.\}\}$, which contains a pair of
values and the code necessary to perform division. The legacy, or
passive packet, is defined as $RUN\{\{\#1/\#2\&\}\},\{22.,7.\}\}$,
which pre-computes the result of the division and transmits the same
information in non-algorithmic form. The argument defined as
$\{\{1,2,3,4\},\{4,3,2,1\}\}$ identifies the links traversed by the
active and passive packets respectively. In this case, the first
packet begins by crossing link one and the second packet begins by
crossing link four. The argument defined as $\{100,100,1000,1000\}$
indicates link capacities for links one, two, three, and four. Thus,
the first packet transmits both code and data that generates the
intended information, while the second packet transmits raw data
only. The result of executing the function below is load and
processing time spent on each link and node for each packet. In Figure
\ref{Kloadgraph}, the load induced by sending the estimate of $\pi$
using {\it AnetSim} in Figure \ref{Kloadgraph} is plotted for each
link. Clearly, the algorithmic representation of the information is
more compact and used less link capacity. In fact, this reinforces the
fact that, by knowing how to compute $\pi$, one could build a more
compact representation; this demonstrates Occam's Razor for a useful
purpose, information compression. This has facilitated study of active
(algorithmic) versus passive transmission of information. For example,
we allow the ratio of data to code to change for the same information
as the packet traverses the network in a manner that optimizes both
link capacity and node processor speed.

\begin{figure*}[ht] 
  \centerline{\psfig{file=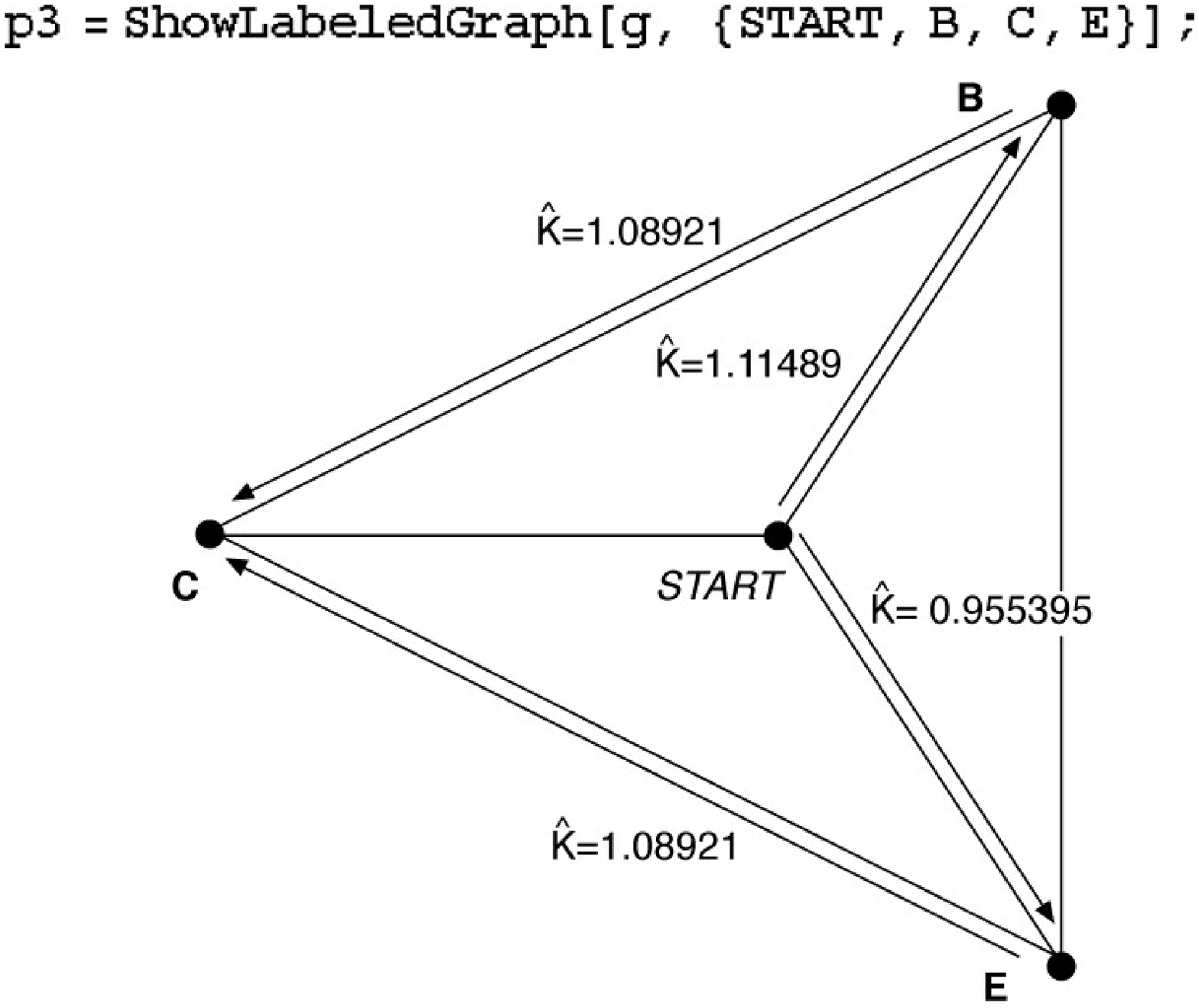,width=5.8in}}
  \caption{\label{SUA}System under Analysis: Components and Topology.}
\end{figure*}

\begin{figure*}[ht]
  \centerline{\psfig{file=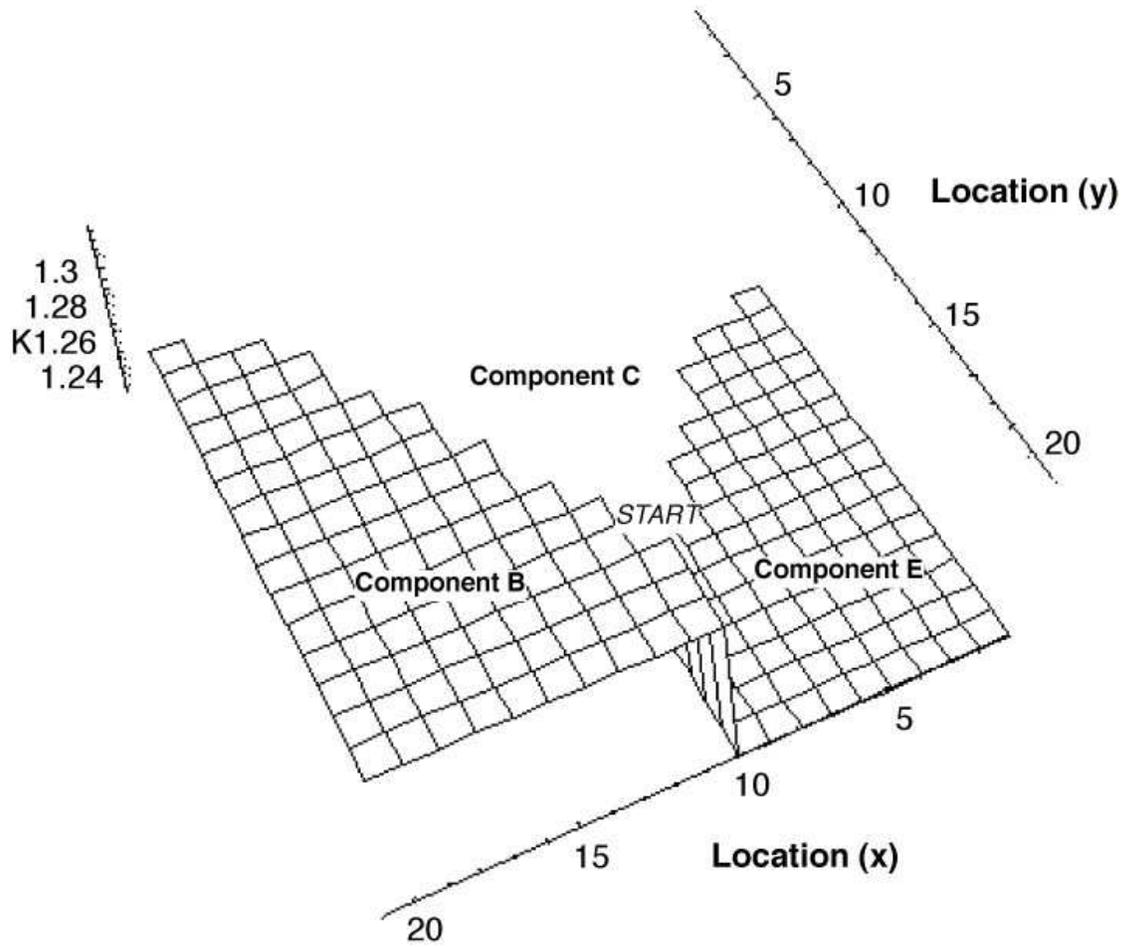,width=5.8in}}
  \caption{\label{Kcontour}K-Map Contour of System in Figures
     \ref{SUA}.}
\end{figure*}

Before continuing with a specific analysis, consider Figure \ref{SUA}
which shows a topological view of components of a sample system under
analysis. The nodes are active application components and the links
are security relationships between the components; the links are
quantified using Kolmogorov Complexity-based vulnerability
analysis. START is a state that is outside the system that represents
the state of the system before it has been penetrated by an attacker.

\begin{figure*}[ht] 
  \centerline{\psfig{file=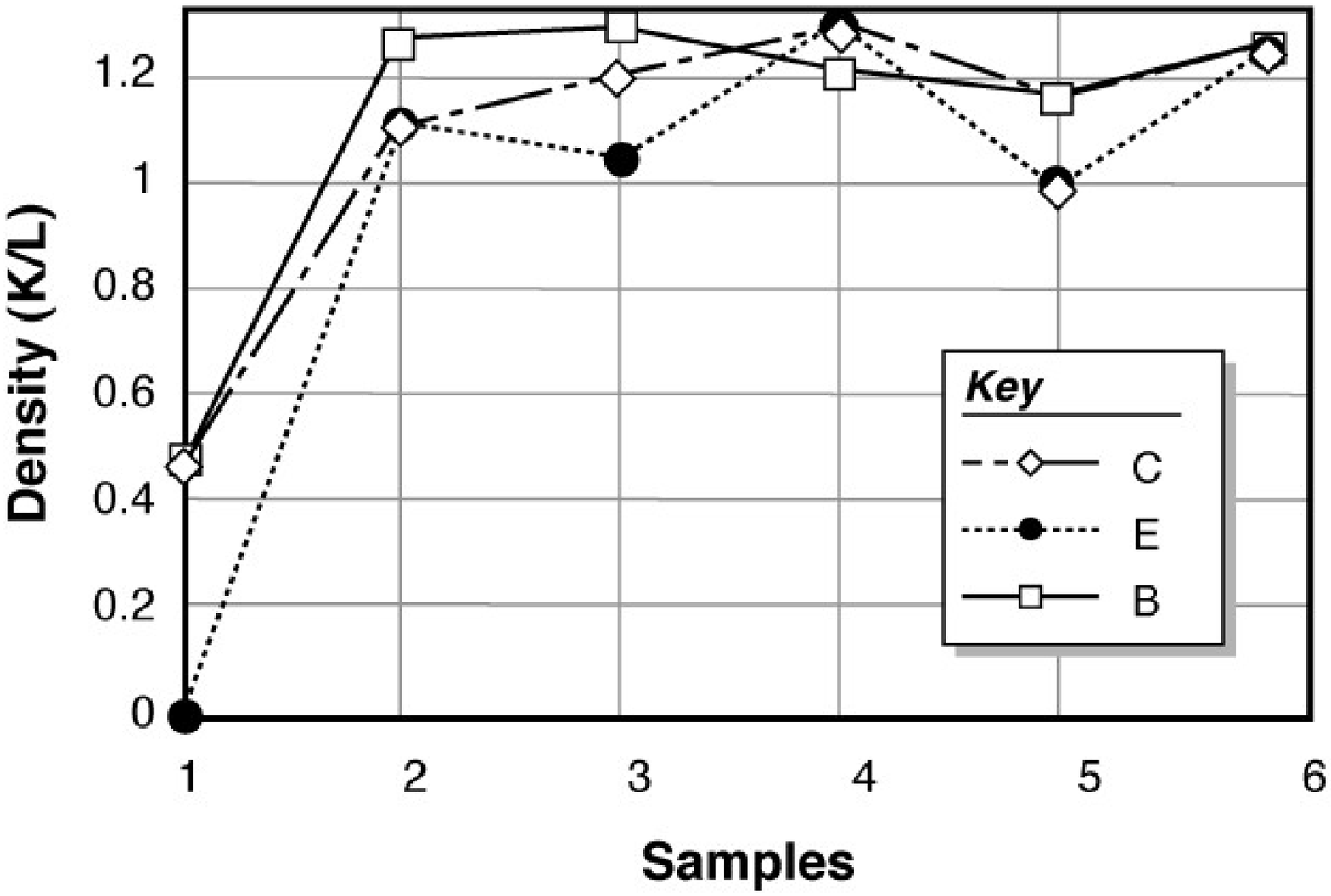,width=5.8in}}
  \caption{\label{compK}Component Complexity for Components B, C, and E.}
\end{figure*}

\SubSection{Complexity Surface: The Kolmogorov Complexity Map}
The General Electric Corporate Research and Development active network
test bed implements complexity probes as part of the active execution
environment. The choice was made to embed the complexity probe in the
execution environment rather than as an active application because it
is necessary to examine the content of active packets before they
reach the execution environment. In the Mathematica simulation, each
component of the active application contains probe-input points
through which bit level input and output is collected. A complexity
estimator based upon the simple inverse compression ratio from
Equation \ref{eq:1} is used to estimate complexity in the density
metric. Figure \ref{compK} graphs results from density estimates taken
of accumulated input and output of three separate components of the
active network application. The graph shows the complexity of
bit-level input and output strings concatenated together. That is,
every input sequence is concatenated with an output sequence and the
density of the sequence is recorded at the bit-level. The input/output
concatenation is generated either for individual components of the
system of for a composition of components. If there is low complexity
in the input/output observation pairs, then it is likely to be easy
for an attacker to understand the system, as in Definition 1. The
X-axis is the number of input and output observations concatenated to
form a single string of bits. From Figure \ref{compK}, it would appear
that Component E is most vulnerable due to its consistently low
complexity while Component B appears to be the least vulnerable due to
its larger complexity. These results make intuitive sense because
Component E simply forwards data without any form of protection while
Component B adds noise to the data. This vulnerability method does not
take into account whether a component reduces or increases complexity;
in other words whether the change was endothermic or exothermic
complexity. These results demonstrate how vulnerabilities are
systematically discovered using complexity; vulnerabilities can be
quantified to a value within the bounds of the complexity measure
error.

\begin{figure}[ht] 
  \centerline{\psfig{file=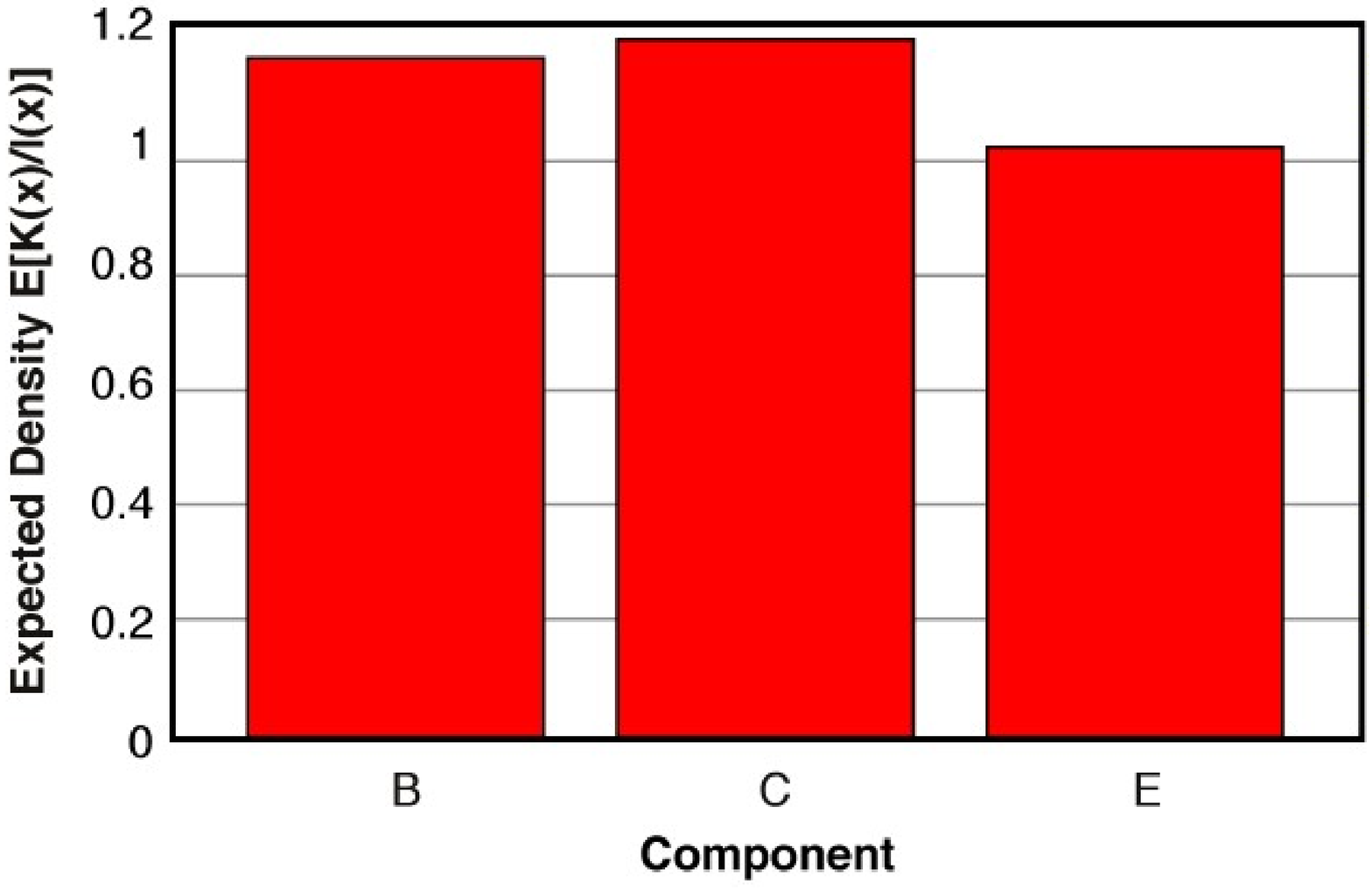,width=3.2in}}
  \caption{\label{uK}Mean Component Complexities for Components B, C, and E.}
\end{figure}

\begin{figure}[ht] 
  \centerline{\psfig{file=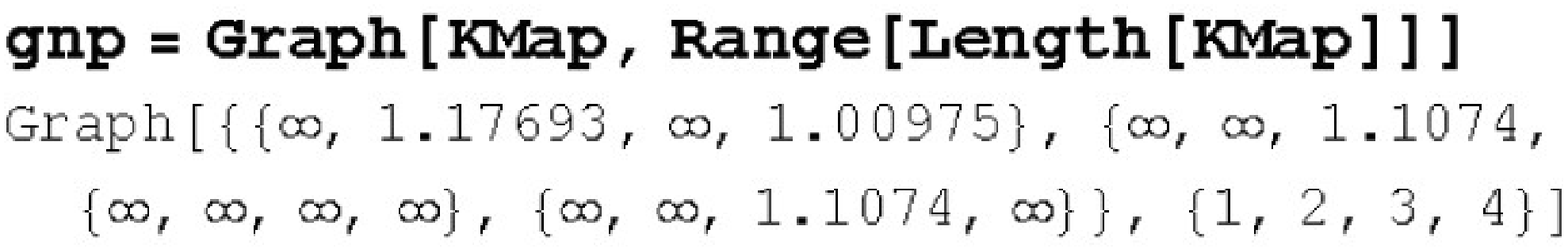,width=3.2in}}
  \caption{\label{Kmap}Kolmogorov Complexity Map (K-Map) Matrix.}
\end{figure}

\begin{figure}[ht] 
  \centerline{\psfig{file=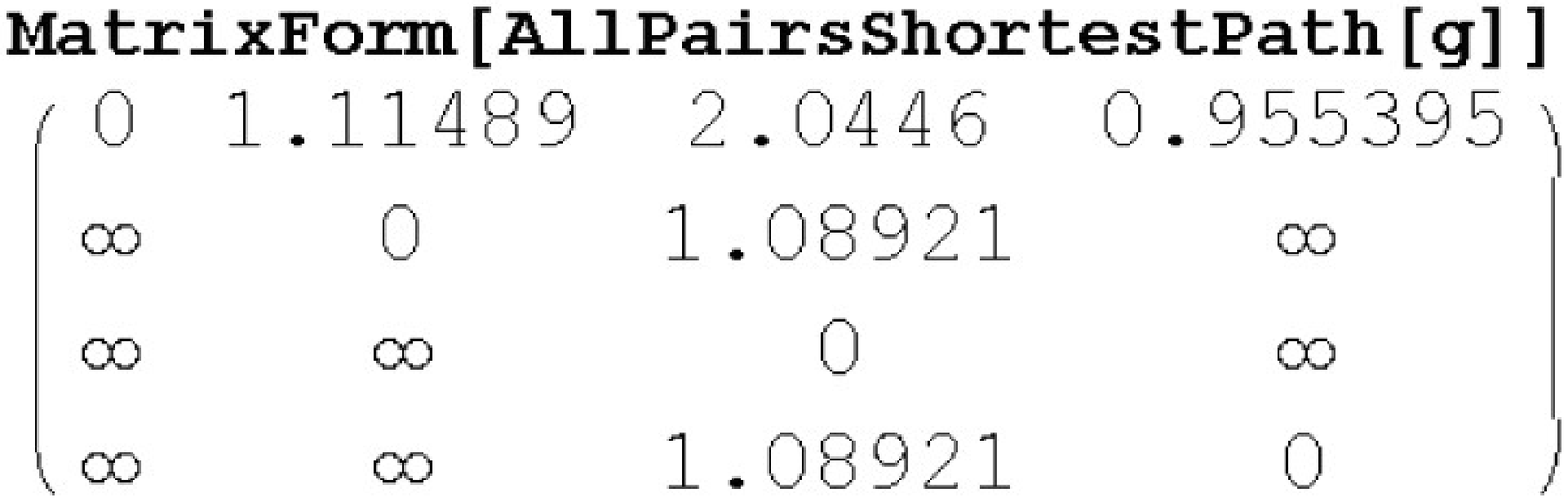,width=3.2in}}
  \caption{\label{minK}Minimum Complexity Paths Matrix.}
\end{figure}

In order to develop the Kolmogorov Complexity Map (K-Map), consider
the topology in more detail. Figure \ref{Kmap} shows the resulting
densities inserted into a Mathematica graph object. The graph object
allows graph theory related analyses to be applied. The directed graph
in Figure \ref{SUA} shows the relationship among the
vulnerabilities. The START state, located in the center of the
topology, represents a location outside the system. In Figure
\ref{minK} a matrix is generated that shows the cost, in terms of
complexity, of traveling from any node to any other node in the K-Map.

\begin{figure}[ht] 
  \centerline{\psfig{file=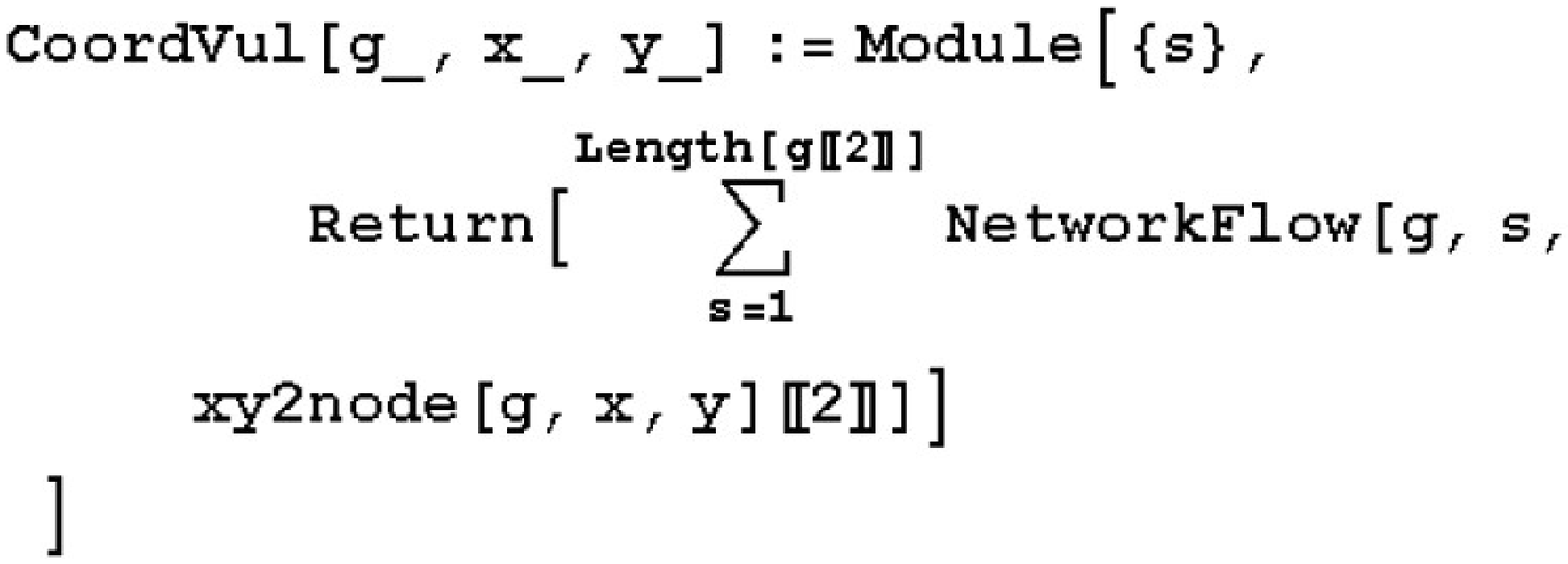,width=3.2in}}
  \caption{\label{iflow}Insecurity Flow Graph.}
\end{figure}

\begin{figure}[ht] 
  \centerline{\psfig{file=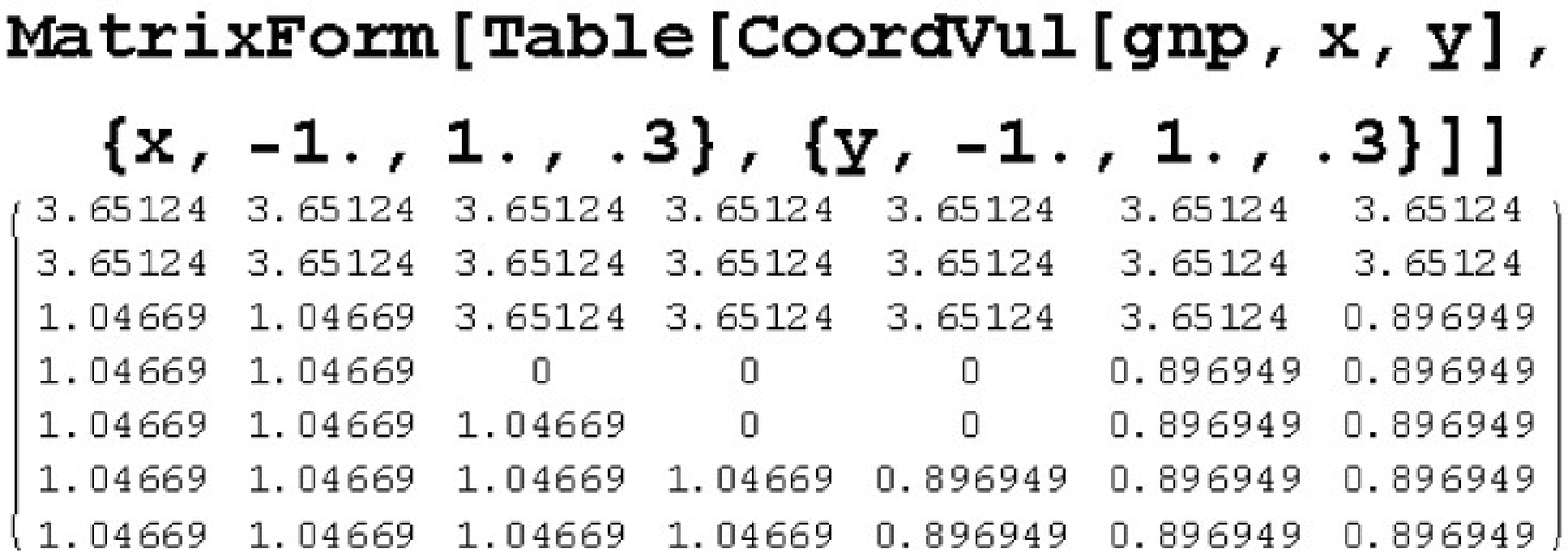,width=3.2in}}
  \caption{\label{fres}Insecurity Flow Results.}
\end{figure}

\begin{figure*}[ht] 
  \centerline{\psfig{file=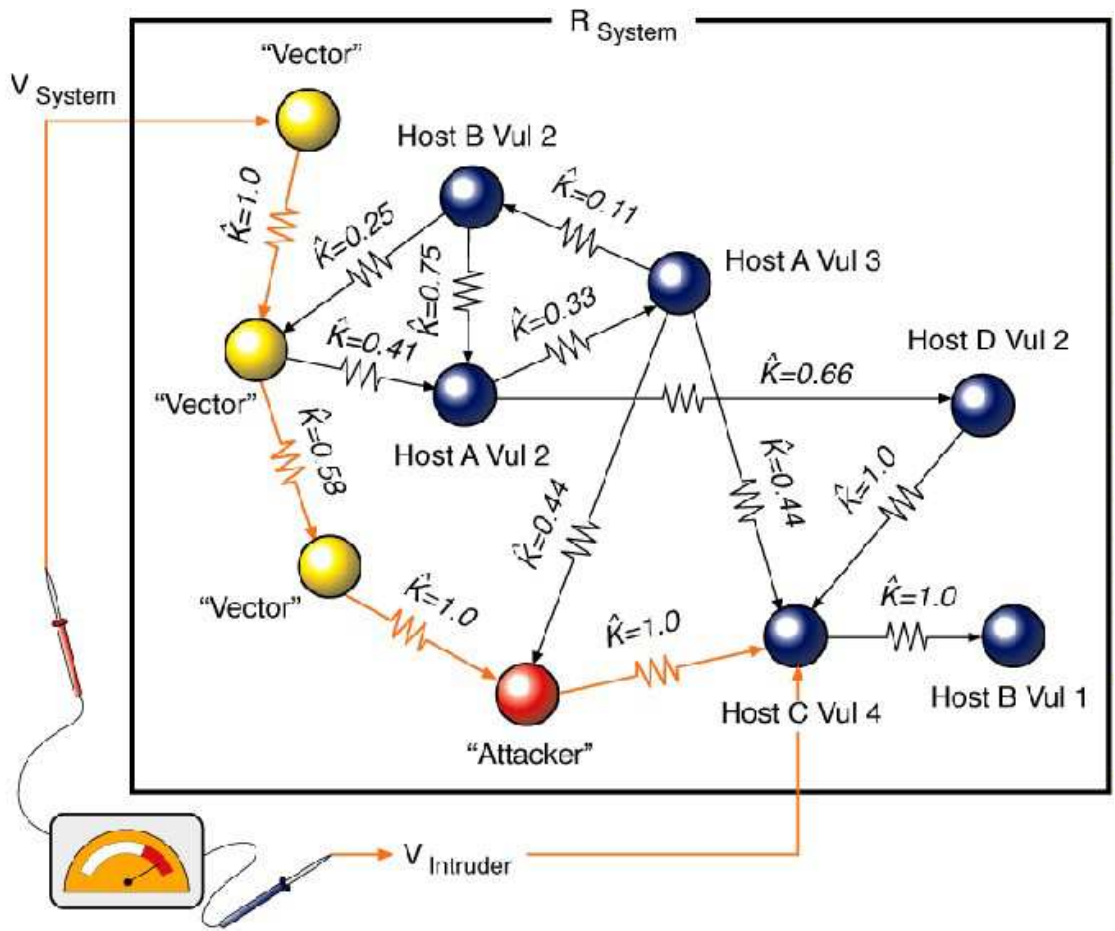,width=5.8in}}
  \caption{\label{iagrid}Grid-Based Representation of Information
     Assurance.}
\end{figure*}

In Figure \ref{iflow} the function {\it CoordVul} computes a maximum
flow through the K-Map graph using the node positions as shown in
Figure \ref{SUA}. Density ($K(x)/l(x)$) from Definition 1, acts as a
resistance, while its inverse acts as conductance, supporting
insecurity flows as illustrated in Figure \ref{iagrid}. The resulting
flow matrix in Figure \ref{fres} shows the maximum flow through each
link. Figure \ref{Kcontour} shows the complexity surface of the
resulting flows. Higher areas correspond to less vulnerable states,
while lower areas correspond to more vulnerable states. Note that in
the following contour maps, areas of infinite height are simply shown
without a surface. By comparing Figure \ref{SUA} and Figure
\ref{Kcontour}, it is apparent that the START state, the infinite
mountain in the center of the topology, is invulnerable, which makes
intuitive sense. State E is the weakest individual component and
lowest area on the right side. Note that while State C cannot be
directly attacked from the START state, it can be attacked via states
B and E, located in the upper and lower right side of the figure
respectively, have a relatively intermediate level of vulnerability.

\begin{figure*}[ht]
  \centerline{\psfig{file=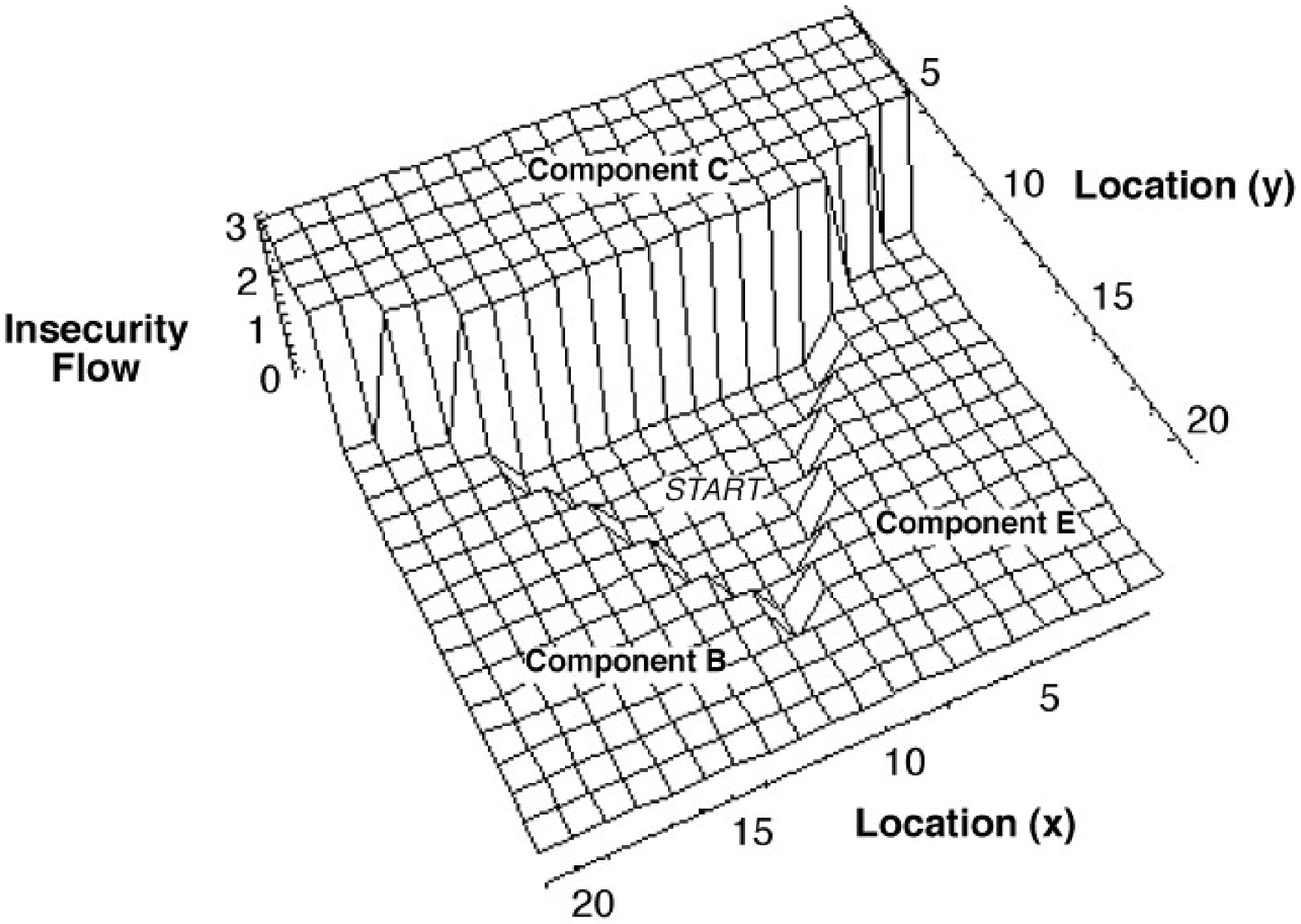,width=5.8in}}
  \caption{\label{Kplot}Insecurity Flow Contour of System in Figure
     \ref{SUA}.} 
\end{figure*}

In the insecurity flow contour shown in Figure \ref{Kplot}, the
density from Definition 1 is resistance and all possible flows from
and to every node are summed to obtain an insecurity level. While Node
C is assigned infinite complexity as shown in Figure \ref{Kcontour},
it actually is the most insecure component given that flows exist from
Nodes B and E.

\Section{Summary}
A Kolmogorov Complexity estimate was used within the Active Virtual
Network Management Prediction framework in order to characterize and
improve system performance. The application in this paper focused on
an active network in which information, algorithmic and static, was
transmitted to support prediction for active network
management. However, the results are ubiquitously applicable to
algorithmic transmission of information in general. Kolmogorov
Complexity was experimentally validated as a theory describing the
relationship between algorithmic compression, complexity, and
prediction accuracy within an active network. Next the relationship
between complexity and vulnerability analysis was proposed. Finally,
this work sets the stage for research into self-composing solutions
based upon Kolmogorov Complexity which will be the focus of the next
phase of this project.

\SubSection{Acknowledgments}
The work discussed in this paper was funded in full by DARPA, under
the auspices of the Fault Tolerant Networks program. Our thanks go to
Doug Maughan, the Active Networks program manager and Scott Shyne, Air
Force Rome Labs for their generous support. 

\bibliographystyle{latex8}
\bibliography{ftn,iw,contrib}

\end{document}